\documentclass[twocolumn,aps,prb,10pt,nofootinbib]{revtex4-2}

\usepackage{amsmath}
\usepackage{amssymb}
\usepackage{wasysym}
\usepackage{graphicx}
\usepackage{hyperref}
\usepackage{dsfont}
\usepackage{newtxtext}
\usepackage[varvw]{newtxmath}
\usepackage{mathtools}
\usepackage[shortlabels]{enumitem}
\usepackage[percent]{overpic}
\usepackage{physics}
\usepackage[normalem]{ulem}

\hypersetup{
    colorlinks,
    linkcolor={blue},
    citecolor={blue},
    urlcolor={blue}
}

\graphicspath{{./}{./figs/}}

\newcommand{\rmd}{\mathrm{d}}

\newcommand{\NCTO}{$\mathrm{Na_2Co_2TeO_6}$}

\DeclareMathOperator{\diag}{diag}

\begin{document}

\title{%
Competing magnetic and spin vestigial orders from continuum field theory%
}

\author{Niccol\`o Francini}
\affiliation{Institut f\"ur Theoretische Physik and W\"urzburg-Dresden Cluster of Excellence ctd.qmat, TU Dresden, 01062 Dresden, Germany}
\author{Lukas Janssen}
\affiliation{Institut f\"ur Theoretische Physik and W\"urzburg-Dresden Cluster of Excellence ctd.qmat, TU Dresden, 01062 Dresden, Germany}

\begin{abstract}
We study magnetic systems featuring competing antiferromagnetic and spin-nematic phases, described by dipolar primary and quadrupolar secondary order parameters. While the antiferromagnetic phase breaks time-reversal and spin-rotational symmetries, only spin-rotational symmetry is broken in the nematic phase, which can be understood as a spin-vestigial phase. We develop a classical continuum field theory with independent vector and tensor fields and use mean-field and renormalization group methods to analyze its phases and finite-temperature transitions. We determine the fixed-point structure using an $\epsilon$ expansion about the upper critical dimension of six and track the resulting fixed points to lower dimensions using a perturbative fixed-dimension renormalization group approach. The multicritical fixed point governing the meeting of the antiferromagnetic, nematic, and paramagnetic phases occurs at imaginary coupling throughout the dimensions accessible to our analysis, indicating a first-order transition through the triple point. Away from the triple point, a direct paramagnetic-to-antiferromagnetic transition can be continuous when the tensor mass is sufficiently large, without an intervening vestigial phase. We argue that it is governed by the cubic universality class in $d=3$ and the Ising universality class in $d=2$ for three-component order parameters. The theory also supports a two-step transition with an intermediate spin-nematic phase. The paramagnetic-to-nematic transition is governed by the four-state Potts universality class in $d=2$ and is first order in $d=3$, while the nematic-to-antiferromagnetic transition is generically expected to be continuous and of Ising type in both dimensions. Our results provide a field-theoretical framework for understanding the finite-temperature transitions observed in the candidate Kitaev material Na$_2$Co$_2$TeO$_6$.
\end{abstract}

\date{\today}

\maketitle

\section{Introduction}
\label{sec:intro}

Vestigial phases are phases of matter in which a composite order develops without the corresponding primary order~\cite{fernandes19}. In the conventional framework, condensation of a primary multicomponent order parameter $\boldsymbol{\phi}$ breaks a set of symmetries, defining the primary phase. In a vestigial or composite phase, by contrast, the primary order parameter remains disordered, $\langle\phi_a\rangle=0$, while composite combinations can acquire a finite expectation value, $\langle\phi_a\phi_b\rangle\neq 0$. These composite operators act as secondary order parameters and characterize the symmetry breaking associated with a subgroup $G$ of the symmetries broken in the primary phase. Vestigial phases can thus be viewed as partially melted versions of the underlying primary phase, in which some, but not all, of its broken symmetries are restored. This mechanism gives rise to a rich variety of intermediate phases and correspondingly complex phase diagrams.

Spin nematicity represents a class of vestigial phases in magnetic systems. Here, the term spin nematic typically refers to a state that breaks spin-rotational symmetry without developing a magnetic moment~\cite{penc2011}. Instead, the order is multipolar, with quadrupolar order providing the simplest example. To illustrate this, consider a multicomponent dipolar order parameter $\boldsymbol{\phi}$, such as the staggered magnetization, which transforms as $\boldsymbol{\phi}\mapsto-\boldsymbol{\phi}$ under time reversal and as $\boldsymbol{\phi}\mapsto R\boldsymbol{\phi}$, with $R\in\mathrm{SO}(N_\phi)$, under spin rotations. In a nematic phase, the dipolar order vanishes, $\langle\phi_a\rangle=0$, while higher moments can acquire finite expectation values. For instance, a nonzero quadrupolar order parameter $\langle\phi_a\phi_b\rangle$ generically breaks spin-rotational symmetry, while preserving time-reversal symmetry~\cite{hecker24,palle26}.

Spin-nematic phases have been discussed in both two- and three-dimensional magnetic systems~\cite{blume69,andreev84,papanicolau88,penc2011,chubukov90,chalker92,shannon06,shannon10,sun23,pohle23,francini25HigherRank,francini2025exactnematic,mishra04,cannas06}. In a previous work, we studied the finite-temperature phase diagram of a two-dimensional classical extended Kitaev-Heisenberg model using extensive Monte Carlo simulations~\cite{francini24vestigial}, motivated by the various transition observed experimentally in the candidate Kitaev material \NCTO~\cite{chen21}. We found that, in a certain region of the phase diagram, the system undergoes two successive transitions upon cooling, with a $\mathbb{Z}_4$ spin-current density wave emerging as a vestigial phase of an underlying triple-$\mathbf{q}$ primary order.
Using a duality transformation~\cite{chaloupka15}, the model can be mapped onto an antiferromagnetic Heisenberg model perturbed by an interaction with cubic symmetry $O_h$, which favors the dual spins to point along the corners of a cube~\cite{krueger23}. In the dual representation, the $\mathbb{Z}_4$ spin-current density wave corresponds to a spin-nematic phase in which the cubic rotational symmetry is spontaneously broken down to a residual $\mathbb{Z}_2$ symmetry, thereby preventing the development of dipolar order.
While these numerical results provide a clear qualitative picture, several questions remain open. Depending on whether an intermediate-temperature vestigial phase is present, the phase diagram exhibits either one or two transitions, whose nature could not be uniquely determined from the numerical simulations because of significant finite-size effects. Moreover, we proposed a renormalization group (RG) flow diagram that qualitatively accounts for the observed phase structure. To our knowledge, however, a continuum field theory that realizes this flow has not yet been established.

\begin{figure*}
\centering
\begin{overpic}[width=\textwidth]{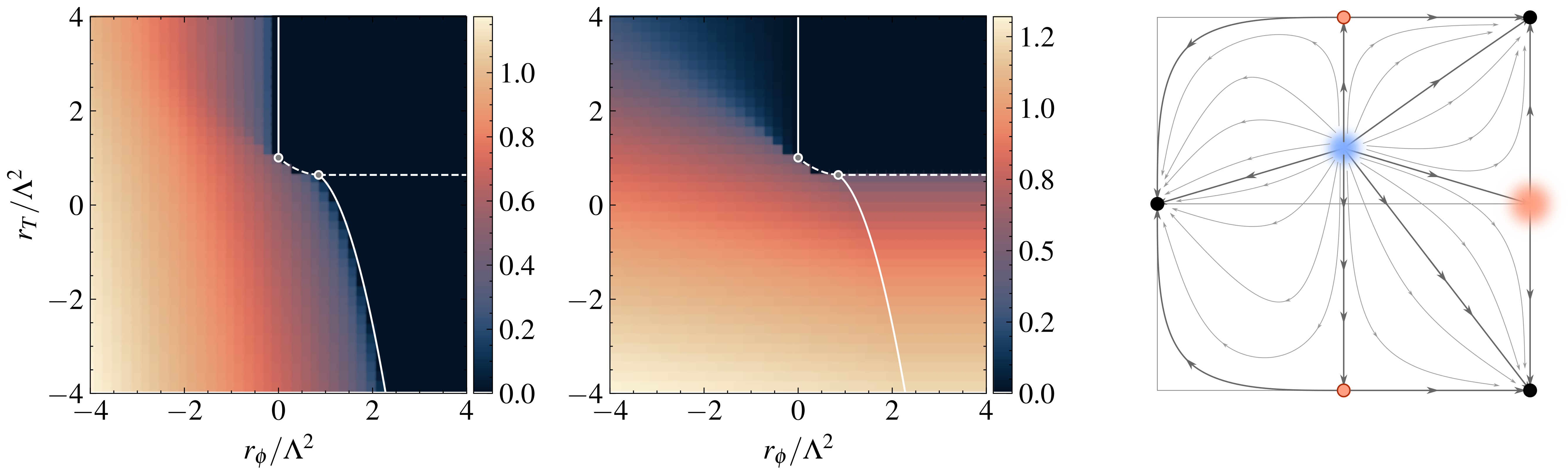}
\put(6,30){(a)}
\put(40,30){(b)}
\put(74,30){(c)}
\put(10,12){AFM}
\put(25,12){\textcolor{white}{SN}}
\put(25,25){\textcolor{white}{PM}}
\put(16,20){\textcolor{white}{B}}
\put(20,19.8){\textcolor{white}{A}} 
\put(43,12){AFM}
\put(58,12){\textcolor{white}{SN}}
\put(58,25){\textcolor{white}{PM}}
\put(49,20){\textcolor{white}{B}}
\put(53,19.8){\textcolor{white}{A}} 
%
\put(27,30){$|\langle\boldsymbol{\phi}\rangle|/\Lambda^\frac{d-2}{2}$}
\put(60,30){$|\langle\boldsymbol{\psi}\rangle|/\Lambda^\frac{d-2}{2}$}
\put(83.5,3){$r_\phi/\Lambda^2$}
\put(70.5,16){\rotatebox{90}{$r_T/\Lambda^2$}}
\put(72,3){$-\infty$}
\put(97,2.5){$\infty$}
\put(72,27.5){$\infty$}
\put(70.9,4){$-\infty$}
\put(81.5,21){MC}
\put(98.5,15){4-Potts}
\put(98.5,28){PM}
\put(98.5,5.5){SN}
\put(74.5,15.5){AFM}
\put(81.4,6){Ising}
\put(81,30.3){$\mathrm{O}(N_\phi)$/cubic}
\end{overpic}
\caption{%
(a)~Antiferromagnetic order parameter $|\langle\boldsymbol{\phi}\rangle|$ from mean-field theory as a function of the mass parameters $r_\phi$ and $r_T$ of the vector and tensor fields, respectively. The interaction parameters are fixed to $(u_\phi\Lambda^{d-4},v_\phi\Lambda^{d-4},u_T\Lambda^{d-4},\omega\Lambda^{(d-6)/2},\lambda\Lambda^{(d-6)/2})=(24,12,36,7.2,1.6)$.
%
(b)~Same as (a), but for the nematic order parameter $|\langle\boldsymbol{\psi}\rangle|$. The phase diagram features three phases: a symmetric paramagnetic (PM) phase, in which the full symmetry group $O_h\simeq S_4\times\mathbb{Z}_2$ is preserved; a quadrupolar spin-nematic (SN) phase, in which the $S_4$ subgroup of $O_h$ is spontaneously broken while the residual $\mathbb{Z}_2$ time reversal symmetry is preserved; and a dipolar antiferromagnetic (AFM) phase, in which both $S_4$ and $\mathbb{Z}_2$ symmetries are spontaneously broken. Solid (dashed) lines indicate continuous (first-order) phase transitions. ``A'' denotes the triple point at which all three phases meet, while ``B'' denotes the point above which the direct paramagnetic-to-antiferromagnetic transition becomes continuous.
(c)~Sketch of the RG flow in the $(r_\phi,r_T)$ plane in $2<d<4$ dimensions. Shaded regions indicate the projections of the multicritical (MC) and four-state Potts (4-Potts) fixed points, which occur at imaginary values of the couplings $\omega$ and $\lambda$. Black dots denote the three real stable fixed points corresponding to the three stable phases, while orange dots denote the two real critical fixed points, which belong to the $\mathrm{O}(N_\phi)$ or cubic universality class, depending on $N_\phi$, and the Ising universality class, respectively.
}
\label{fig:mean-field-PD}
\end{figure*}
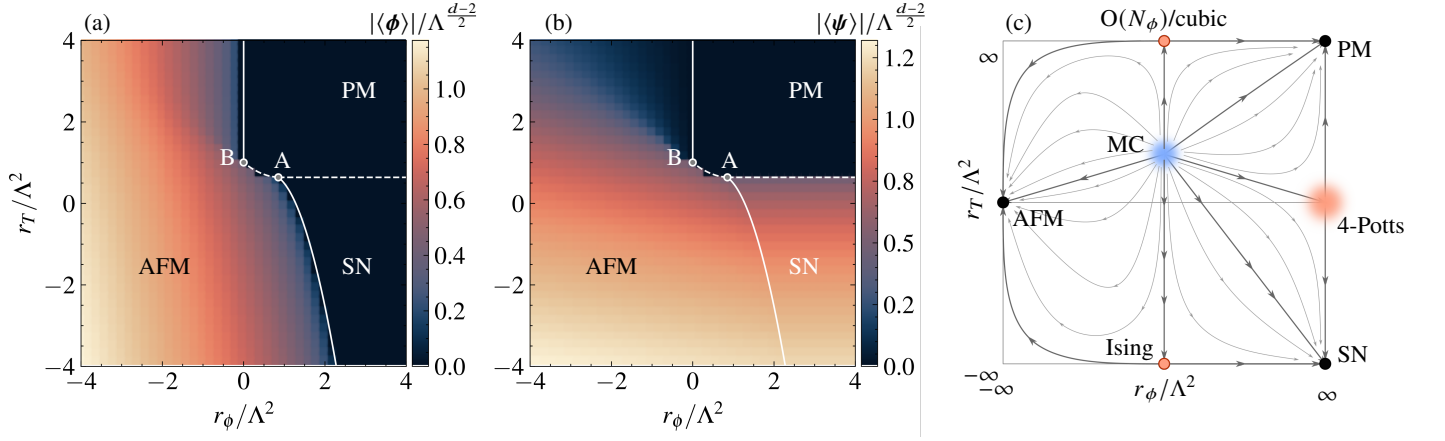

In this work, we develop a minimal continuum field theory to identify the possible critical scenarios arising from competition between primary dipolar and secondary quadrupolar order. We consider a vector order parameter $\boldsymbol{\phi}$ describing a dipolar antiferromagnetic phase and a tensor order parameter $T$ characterizing a spin-nematic phase. The latter is related to the primary order through $T_{ab}\sim\sum_{cd}t_{abcd}\phi_c\phi_d$, where $t_{abcd}$ specifies which combinations of components of the primary order parameter contribute to the nematic order. The two fields interact through the simplest symmetry-allowed coupling, $\boldsymbol{\phi}^\top T\boldsymbol{\phi}$, giving rise to the three expected phases: paramagnetic, nematic, and antiferromagnetic. Our results are summarized in Fig.~\ref{fig:mean-field-PD}. Figures~\ref{fig:mean-field-PD}(a) and (b) show the mean-field order parameters $\langle\boldsymbol{\phi}\rangle$ and $\langle\boldsymbol{\psi}\rangle$, respectively, as functions of the tuning parameters $r_\phi$ and $r_T$, where $\boldsymbol{\psi}$ denotes the irreducible components of the tensor field $T$.
Our one-loop momentum-shell RG analysis incorporates fluctuations of both fields and yields the schematic flow diagram in the $(r_\phi,r_T)$ plane shown in Fig.~\ref{fig:mean-field-PD}(c).
In particular, these results suggest that
(1)~the direct transition between the paramagnetic and multicomponent antiferromagnetic phases, obtained by varying $r_\phi$ at sufficiently large fixed $r_T$, remains direct and continuous beyond mean-field theory when fluctuations are included, contrary to generic expectations~\cite{golubovic88, fernandes12, fernandes19, hecker23};
(2)~the triple point at which the paramagnetic, vestigial spin-nematic, and primary antiferromagnetic phases meet is first order. This follows from the fact that the corresponding multicritical fixed point [MC in Fig.~\ref{fig:mean-field-PD}(c)] occurs at imaginary values of some of the couplings.

The paper is organized as follows. In Sec.~\ref{sec:model}, we introduce the field theory for the vector and tensor degrees of freedom, including the simplest symmetry-allowed vector-tensor interaction. Section~\ref{sec:mean-field-theory} presents the mean-field analysis, where fluctuations are neglected in a Landau-theory framework. In Sec.~\ref{sec:renormalization-group}, we incorporate fluctuations using a one-loop momentum-shell RG approach. We analyze the critical behavior within an $\epsilon$ expansion about six dimensions, as well as using a perturbative analysis in fixed dimensions below six. The validity of the perturbative approach is assessed through comparison with nonperturbative results in low dimensions for selected limits of the phase diagram. We conclude in Sec.~\ref{sec:conclusion}.

\section{Model}
\label{sec:model}

We consider a classical magnetic system with cubic spin-rotational symmetry, as may arise from spin-orbit coupling in a cubic local crystal environment, on a $d$-dimensional lattice. The corresponding continuum field theory is formulated in terms of a vector field $\boldsymbol{\phi}$ describing the primary antiferromagnetic order and a symmetric, traceless rank-two tensor~$T$~\cite{herbut16,pelissetto18,bonati25} describing the vestigial nematic order.
This setup is motivated by the competition between the $\mathbb Z_4$ spin-current density wave and triple-$\mathbf q$ antiferromagnetic order found in the extended Kitaev-Heisenberg model on the honeycomb lattice~\cite{francini24vestigial}, but may also be relevant to possible higher-dimensional realizations of Kitaev-Heisenberg models~\cite{krueger20}.

A common approach to studying vestigial phases is based on a $1/N$ expansion. In this framework, the Hamiltonian contains quartic selfinteractions of the primary vector field $\boldsymbol{\phi}$ of the form $\phi_i\phi_j\phi_k\phi_l$. These interactions can be decoupled by introducing Hubbard-Stratonovich fields $\psi_a=\boldsymbol{\phi}^\top M_a\boldsymbol{\phi}$, where $M_a$ are symmetric matrices. Here, the fields $\psi_a$ correspond to the irreducible components of the tensor field $T$. After integrating out $\boldsymbol{\phi}$, one obtains an effective action for the composite fields $\psi_a$, which can then be treated within a saddle-point approximation that becomes exact in the limit $N\rightarrow\infty$~\cite{hecker24,obrien26vestigialnematicorderzero}.
This procedure, however, is subject to a Fierz ambiguity. Because the composite fields are not independent, different decompositions of the quartic interactions into Hubbard-Stratonovich channels can leave the original Hamiltonian unchanged while modifying the corresponding saddle-point equations. Consequently, the resulting saddle-point solution generally depends on the choice of decomposition and may lead to different conclusions about which vestigial channel condenses~\cite{jaeckel03,herbut16,palle26}.
A procedure for resolving this Fierz ambiguity within the $1/N$ expansion was recently proposed in Ref.~\cite{palle26}.

In this work, we follow a different approach. Rather than integrating out the vector degrees of freedom, we retain both the vector and tensor degrees of freedom as independent coarse-grained fields and treat their fluctuations on equal footing. This keeps both critical sectors explicit and provides a natural framework for describing their competition. Importantly, the tensor field is introduced directly as an independent order-parameter field, rather than through a Hubbard-Stratonovich decomposition of the quartic vector interactions. In the limits where one field is sufficiently massive, it can be integrated out, recovering an effective theory for the remaining critical field.

\subsection{Vector sector}

\begin{figure}
\centering
\begin{overpic}[width=0.8\linewidth]{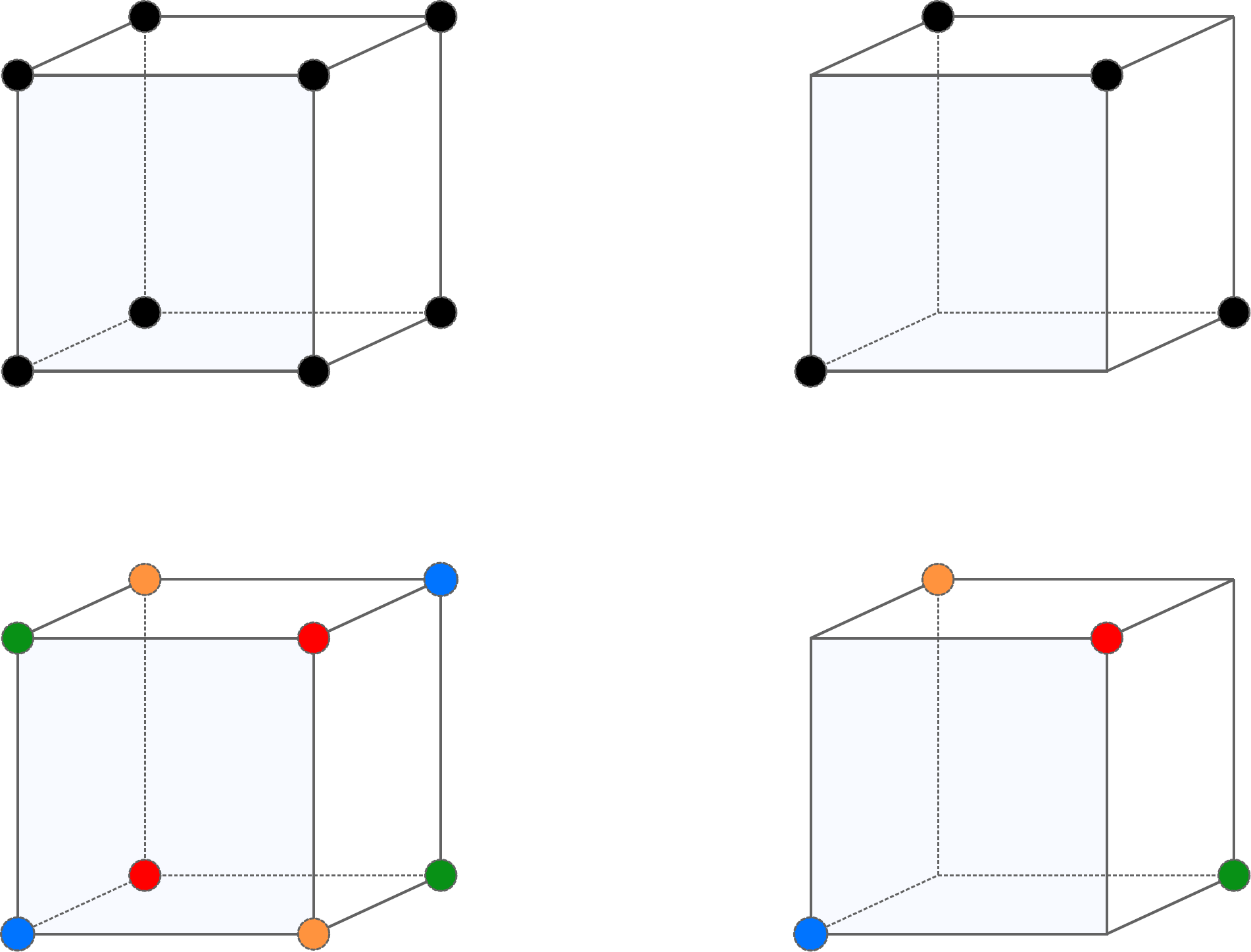}
\put(10,42){$\phi_x$}
\put(30,45){$\phi_y$}
\put(-5,56){$\phi_z$}
\put(-8,70){(a)}
\put(5,78){$v_\phi>0$, $\lambda=0$}
\put(74,42){$\psi_1$}
\put(94,45){$\psi_2$}
\put(59,56){$\psi_3$}
\put(56,70){(b)}
\put(70,78){$\omega>0$, $\lambda=0$}
\put(10,-3){$\phi_x$}
\put(30,0){$\phi_y$}
\put(-5,11){$\phi_z$}
\put(-8,25){(c)}
\put(5,33){$v_\phi,\omega,\lambda>0$}
\put(74,-3){$\psi_1$}
\put(94,0){$\psi_2$}
\put(59,11){$\psi_3$}
\put(56,25){(d)}
\put(70,33){$v_\phi,\omega,\lambda>0$}
\end{overpic}
\caption{%
Ground-state configurations of the vector field $\boldsymbol{\phi}$ (left column) and tensor field $\boldsymbol{\psi}$ (right column) for vanishing ($\lambda=0$, top row) and finite ($\lambda\neq0$, bottom row) vector-tensor coupling.
(a)~For $v_\phi>0$ and sufficiently negative $r_\phi$, $\boldsymbol{\phi}$ points along one of the eight body diagonals, indicated by the black dots at the vertices of the cube.
(b)~For $\omega>0$ and sufficiently negative $r_T$, $\boldsymbol{\psi}$ points along one of the four body diagonals satisfying $\psi_1\psi_2\psi_3<0$, indicated by the black dots at the vertices of the cube.
(c,d)~A finite $\lambda>0$ couples the configurations of $\boldsymbol{\phi}$ and $\boldsymbol{\psi}$, as indicated by the differently colored dots. In particular, when $\boldsymbol{\psi}$ orders along one of the four body diagonals satisfying $\psi_1\psi_2\psi_3<0$, it lifts the degeneracy of the $\boldsymbol{\phi}$ configurations, reducing the number of degenerate states from eight to two.
}
\label{fig:minima-sketches}
\end{figure}
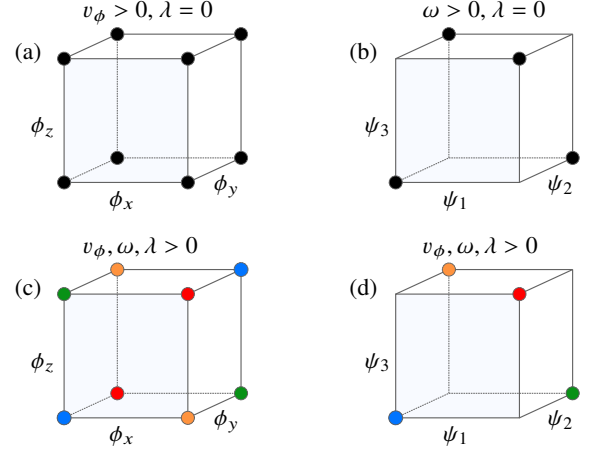

In the presence of cubic symmetry $O_h$, the most general Hamiltonian in the vector sector up to quartic order in the field and quadratic order in derivative is~\cite{pelissetto02}
\begin{align}
\begin{split}
\mathcal{H}_\phi & = \int \rmd^d x \,\Biggl [ \frac{1}{2} \sum_{a=1}^{N_\phi} (\nabla \phi_a)^2 +\frac{r_\phi}{2}\sum_{a=1}^{N_\phi} \phi_a^2 
\\ & \quad
+ \frac{1}{4!}\sum_{a,b=1}^{N_\phi}( u_\phi + v_\phi\delta_{ab} )\phi_a^2\phi_b^2 + \mathcal O(\phi^6) \Biggr]\,,
\label{eq:vector-part}
\end{split}
\end{align}
where $\boldsymbol{\phi} = (\phi_a)$ is a real vector field with $N_\phi$ components, representing the antiferromagnetic order parameter. We specialize to $N_\phi=3$ in the calculations below, while keeping $N_\phi$ general where possible to connect with general results in the literature.
When $d=N_\phi$, so that spatial and internal indices can be identified under the cubic symmetry, an additional kinetic term, $\sum_a(\partial_a\phi_a)^2$, is also allowed. We neglect this term as it is expected to be irrelevant~\cite{pelissetto02,janssen15}.
The coupling $v_\phi$ parametrizes the cubic anisotropy. For $v_\phi>0$, the energy is minimized when $\boldsymbol{\phi}$ points along one of the eight body diagonals, $\boldsymbol{\phi}\propto [111]$ or symmetry-equivalent directions, whereas for $v_\phi<0$ favors $\boldsymbol{\phi}$ to point along one of the six cubic axes, $\boldsymbol{\phi}\propto [001]$ or symmetry-equivalent directions. We focus on $v_\phi>0$, a regime that can be realized, for example, in the extended Kitaev-Heisenberg model on the honeycomb lattice with triple-$\mathbf q$ antiferromagnetic primary order at low temperatures~\cite{krueger23,francini24vestigial}. The continuum theory can also describe potential higher-dimensional realizations, such as those that may occur on hyperhoneycomb or stripyhoneycomb lattices~\cite{krueger20}.
The case $v_\phi < 0$ can be realized in the extended Kitaev-Heisenberg model on the honeycomb lattice in the regime where zigzag antiferromagnetic order is stabilized at low temperatures~\cite{francini24vestigial}.
It may also be relevant for the spin-nematic phase recently observed in nearest-neighbor spin models on the three-dimensional pyrochlore lattice~\cite{francini2025exactnematic}.
The vector-field mass $r_\phi$ serves as a tuning parameter for the transition. For large positive $r_\phi$, the system is in the symmetric phase with $\boldsymbol{\phi}=0$, whereas for sufficiently negative $r_\phi$ the cubic symmetry is spontaneously broken and $\boldsymbol{\phi}$ selects one of the eight equivalent minima with $\boldsymbol{\phi}\propto[111]$ or a symmetry-equivalent direction, as illustrated in Fig.~\ref{fig:minima-sketches}(a).

The Hamiltonian in Eq.~\eqref{eq:vector-part} is invariant under the octahedral group $O_h\simeq S_4\times\mathbb{Z}_2$. The $S_4$ subgroup consists of the 24 proper rotations that leave the cube invariant, while the $\mathbb{Z}_2$ factor acts as inversion of the vector field. For a magnetic order parameter, this latter transformation is also realized by time reversal $\tau$. The vector field therefore transforms under proper cubic rotations as
\begin{align}
\label{eq:Oh-vector-transform-1}
S_4: \quad \boldsymbol{\phi}&\mapsto \boldsymbol{\phi}'=R\boldsymbol{\phi}, \quad
\text{where } R\in O_h \text{ with } \det R=1,
\end{align}
while it changes sign under time reversal,
\begin{align}
\label{eq:Oh-vector-transform-2}
\tau: \quad \boldsymbol{\phi}&\mapsto\boldsymbol{\phi}'=-\boldsymbol{\phi}\,.
\end{align}

%

\subsection{Triplet tensor sector}

In the tensor sector, the most general Hamiltonian up to quartic order in the field and quadratic order in derivatives is~\cite{janssen15,pelissetto18,hecker24,bonati25}
\begin{equation}
\label{eq:raw-nematic-part}
\begin{split}
\mathcal{H}_T &= \int \rmd^d x \,\biggl\{\frac14 \operatorname{Tr}[(\nabla T)^2] + \frac{{r}_T}{4} \operatorname{Tr}(T^2) + \frac{\omega}{3!} \operatorname{Tr}(T^3)
\\ &\quad
+\frac{u_T}{4 \cdot 4!} \left[\operatorname{Tr}(T^2)\right]^2 + \frac{v_T}{4 \cdot 4!} \operatorname{Tr} (T^4) + \mathcal O(T^5) \biggr\}\,,
\end{split}
\end{equation}
where $T$ is a real, symmetric, traceless rank-two tensor with $\frac12 N_T (N_T+1)-1$ independent components. The parameter $r_T$ is the tensor-field mass, while $\omega$, $u_T$, and $v_T$ are self-interaction couplings. In particular, the cubic term $\operatorname{Tr}(T^3)$ is allowed by symmetry and is therefore generically present. The tensor field can be viewed as a quadrupolar order parameter, in contrast to the dipolar vector field $\boldsymbol{\phi}$.
In the calculations below, we specialize to $N_T=3$, while keeping $N_T$ general where possible. The independent components of $T$ can be parametrized using a basis ${\Lambda_a}$ of real symmetric traceless $N_T\times N_T$ matrices~\cite{janssen15} as $T=\sum_a\psi_a\Lambda_a$, where $a=1,\dots,\frac12 N_T(N_T+1)-1$. For $N_T=3$, this gives five independent components $\psi_a$, with the real symmetric Gell-Mann matrices forming a basis of the tensor space. Cubic symmetry does not, however, require these five components to be degenerate in general.

The tensor Hamiltonian in Eq.~\eqref{eq:raw-nematic-part} features the same $O_h \simeq S_4 \times \mathbb Z_2$ symmetry as the vector Hamiltonian in Eq.~\eqref{eq:vector-part}. Under a proper cubic rotation, the tensor field transforms as
\begin{equation}
S_4: \quad T\mapsto T'=R T R^\top,\quad
\text{where } R\in O_h \text{ with } \det R=1,
\end{equation}
while it remains invariant under time reversal,
\begin{equation}
\tau: \quad T\mapsto T'=T
\end{equation}
Under the cubic symmetry group $O_h$, the five-dimensional tensor space decomposes into the two irreducible representations $E_g$ and $T_{2g}$, which we refer to as the doublet and triplet~\cite{hecker24}. The symmetry-related minima of both sectors correspond to uniaxial nematic order: the two-dimensional $E_g$ doublet has three symmetry-related energy minima corresponding to directors oriented along the three cubic axes, while the three-dimensional $T_{2g}$ triplet has four symmetry-related minima corresponding to directors oriented along the four cubic body diagonals, as illustrated in Fig.~\ref{fig:minima-sketches}(b).

In our work, we focus on the triplet component of the nematic tensor, relevant for the competition between the $\mathbb Z_4$ spin-current density wave and triple-$\mathbf q$ antiferromagnetic order found in the extended Kitaev-Heisenberg model on the honeycomb lattice~\cite{francini24vestigial}. It is expressed as 
\begin{align}
\label{eq:tensor-components}
T=\sum_{a=1}^3 \psi_a \Lambda_a\,,
\end{align}
where
\begin{equation}
\label{eq:real-gell-mann}
    \Lambda_1=\begin{pmatrix}
        0 & 0 & 0 \\ 0 & 0 & 1 \\ 0 & 1 & 0 
    \end{pmatrix},\quad  
    \Lambda_2=\begin{pmatrix}
        0 & 0 & 1 \\ 0 & 0 & 0 \\ 1 & 0 & 0 
    \end{pmatrix},\quad
    \Lambda_3=\begin{pmatrix}
        0 & 1 & 0 \\ 1 & 0 & 0 \\ 0 & 0 & 0 
    \end{pmatrix}
\end{equation}
are the three off-diagonal real Gell-Mann matrices.
For $N_T = 3$, the two quartic invariants in Eq.~\eqref{eq:raw-nematic-part} are not independent, since
\begin{align}
\operatorname{Tr}(T^4) = \frac12 \left[\operatorname{Tr}(T^2)\right]^2
\end{align}
for $3\times 3$ traceless symmetric matrices $T$.
The tensor Hamiltonian $\mathcal H_T$ in Eq.~\eqref{eq:raw-nematic-part} can be recast in terms of the three irreducible $T_{2g}$ components $\psi_a$ of the tensor field $T$ as
\begin{equation}
\label{eq:nematic-part}
\begin{split}
\mathcal{H}_T&=\int \rmd^d x \,\biggl[\frac{1}{2}\sum_{a=1}^3(\nabla\psi_a)^2 +\frac{r_T}{2}\sum_{a=1}^3\psi_a^2
\\ &\quad
+ \frac{\omega}{3!}\sum_{a,b,c=1}^3 |\epsilon^{abc}|\psi_a\psi_b\psi_c 
+\frac{u_T}{4!}\sum_{a,b=1}^3\psi_a^2\psi_b^2 
\\ &\quad
+ \mathcal O(\psi^5)\biggr]\,,
\end{split}
\end{equation}
where $\epsilon^{abc}$ is the Levi-Civita symbol and we have absorbed the combination $u_T+\frac12 v_T$ into a redefined quartic coupling, $u_T+\frac12v_T\mapsto u_T$. For $\omega>0$, the cubic invariant is minimized when $\psi_1\psi_2\psi_3<0$. Thus, for sufficiently negative $r_T$, the four-state symmetry is spontaneously broken by selecting one of four degenerate minima with $\psi_1\psi_2\psi_3<0$. These minima correspond to uniaxial nematic order with the director oriented along one of the four cubic body diagonals, as illustrated in Fig.~\ref{fig:minima-sketches}(b). The Hamiltonian in Eq.~\eqref{eq:nematic-part} is equivalent to the four-state Potts model~\cite{zia75,wu82,hecker24}.

\subsection{Full model}

To couple the vector and tensor sectors, we consider the simplest symmetry-allowed interaction,
\begin{equation}
    \label{eq:interacting-term}
    \mathcal{H}_{\phi T} = \frac{\lambda}{2} \int \rmd^dx \,\boldsymbol{\phi}^\top T\boldsymbol{\phi}\,.
\end{equation}
Restricting to the $T_{2g}$ triplet and decomposing $T$ into its irreducible components $\psi_a$, $a=1,2,3$ [Eq.~\eqref{eq:tensor-components}], the interaction takes the form
\begin{equation}
    \label{eq:simplified-interacting-term}
    \mathcal{H}_{\phi T}= \frac{\lambda}{2} \int \rmd^dx \,\sum_{a,b,c=1}^3|\epsilon^{abc}|\psi_a\phi_b\phi_c\,.
\end{equation}

The full Hamiltonian considered in this work is
\begin{equation}
    \label{eq:full-Hamiltonian}
    \mathcal{H}=\mathcal{H}_{\phi}+\mathcal{H}_T+\mathcal{H}_{\phi T}\,.
\end{equation}
The vector-tensor interaction couples the two order parameters and hard-wires the full model to the case $N_\phi=N_T=3$. In particular, when the tensor field $\boldsymbol{\psi}$ orders into one of its four symmetry-related states, it acts as a quadratic anisotropy for the vector field $\boldsymbol{\phi}$. This anisotropy selects one of the cubic body diagonals in $\boldsymbol{\phi}$ space, reducing the eightfold degeneracy of the vector field to two symmetry-related states, as illustrated in Figs.~\ref{fig:minima-sketches}(c,d). Thus, if nematic order develops before antiferromagnetic order, the remaining vector degrees of freedom are effectively described by a $\mathbb{Z}_2$-symmetric Ising model; see also Appendix~\ref{app:rt-minus-infty}.
Conversely, when the vector field orders, the bilinear $\sum_{b,c}|\epsilon^{abc}|\langle\phi_b\phi_c\rangle$ acts as conjugate fields for the nematic order parameter $\psi_a$. Consequently, the antiferromagnetic phase necessarily has nonzero tensor order, lifting the fourfold degeneracy of the nematic states and leaving a unique minimum.
Thus, the interaction in Eq.~\eqref{eq:interacting-term} allows for three distinct phases: a symmetric paramagnetic phase with $\langle\boldsymbol{\phi}\rangle=0$ and $\langle\boldsymbol{\psi}\rangle=0$, a spin-nematic phase with $\langle\boldsymbol{\phi}\rangle=0$ and $\langle\boldsymbol{\psi}\rangle\neq0$, and an antiferromagnetic phase with $\langle\boldsymbol{\phi}\rangle\neq0$ and $\langle\boldsymbol{\psi}\rangle\neq0$.

To illustrate the connection between $\boldsymbol{\phi}$ and $\boldsymbol{\psi}$ further, it is useful to temporarily neglect the self-interactions $\mathcal O(\phi^4)$, $\mathcal O(T^3)$, and $\mathcal O(T^4)$ in the vector and tensor sectors. In this limit, the equation of motion for the tensor field relates the expectation values of $\psi_a$ to those of the vector bilinears $\phi_b\phi_c$ as
\begin{equation}
    \label{eq:secondary-OP-eom}
    \langle\psi_a\rangle\approx-\frac{\lambda}{2r_T}\sum_{b,c=1}^3 |\epsilon^{abc}| \langle \phi_b\phi_c \rangle\,.
\end{equation}
In terms of the original tensor field $T$, this relation reads
\begin{equation}
    \label{eq:secondary-OP-tensor}
    \langle T \rangle \approx-\frac{\lambda}{r_T}\begin{pmatrix}
        0 & \langle\phi_x\phi_y\rangle & \langle\phi_x\phi_z\rangle \\
        \langle\phi_x\phi_y\rangle & 0 & \langle\phi_y\phi_z\rangle \\
        \langle\phi_x\phi_z\rangle & \langle\phi_y\phi_z\rangle & 0
    \end{pmatrix}\,.
\end{equation}
Thus, $T$ is explicitly identified as a secondary order parameter constructed from bilinears of the primary field, providing a direct connection to the vestigial spin-nematic order discussed in the introduction. We emphasize, however, that neither $\boldsymbol{\phi}$ nor $\boldsymbol{\psi}$ is integrated out in what follows. Instead, we retain both as independent coarse-grained fields and treat their fluctuations on equal footing.

\section{Mean-field theory}
\label{sec:mean-field-theory}

\begin{figure}[tb!]
\centering
\begin{overpic}[width=\columnwidth]{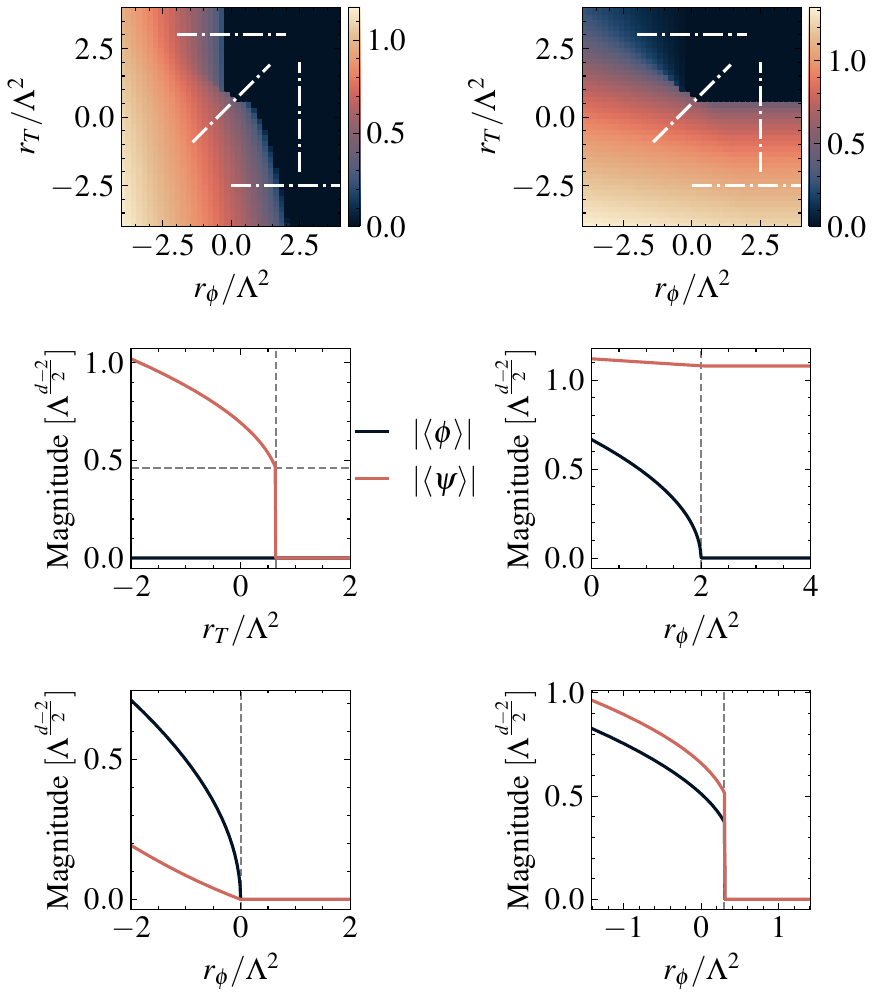}
\put(0,98){(a)}
\put(47,98){(b)}
\put(0,68){(c)}
\put(47,68){(d)}
\put(0,34){(e)}
\put(47,34){(f)}
\put(31,91){\textcolor{white}{1}}
\put(20,80){\textcolor{white}{2}}
\put(15,96){\textcolor{white}{3}}
\put(18,87){\textcolor{white}{4}}
\put(30,101){\footnotesize{$|\langle\boldsymbol{\phi} \rangle|/\Lambda^{\frac{d-2}{2}}$}}
\put(77,91){\textcolor{white}{1}}
\put(66,80){\textcolor{white}{2}}
\put(61,96){\textcolor{white}{3}}
\put(64,87){\textcolor{white}{4}}
\put(76,101){\footnotesize{$|\langle\boldsymbol{\psi} \rangle|/\Lambda^{\frac{d-2}{2}}$}}
\put(13,66){Path 1}
\put(59,66){Path 2}
\put(13,31.5){Path 3}
\put(59,31.5){Path 4}
\put(16,46){SN}
\put(29,46){PM}
\put(61,46){AFM}
\put(74,46){SN}
\put(15,15){AFM}
\put(28,15){PM}
\put(61,15){AFM}
\put(74,15){PM}
\end{overpic}
\caption{%
(a)~Antiferromagnetic order parameter $|\langle\boldsymbol{\phi}\rangle|$ from mean-field theory as a function of the vector and tensor masses $r_\phi$ and $r_T$, respectively. The interaction parameters are fixed as in Fig.~\ref{fig:mean-field-PD}(a,b). Dash-dotted lines indicate the cuts shown in (c)--(f).
(b)~Same as (a), but for the nematic order parameter $|\langle\boldsymbol{\psi}\rangle|$.
(c)Antiferromagnetic order parameter $|\langle\boldsymbol{\phi}\rangle|$ (black) and nematic order parameter $|\langle\boldsymbol{\psi}\rangle|$ (orange) along Path~1, as functions of $r_T$ at fixed $r_\phi/\Lambda^2=2.5$, across the paramagnetic-to-nematic transition.
(d)~Same as (c), but along Path~2, as functions of $r_\phi$ at fixed $r_T/\Lambda^2=-2.5$, across the nematic-to-antiferromagnetic transition.
(e)~Same as (c), but along Path~3, as functions of $r_\phi$ at fixed $r_T/\Lambda^2=3$, across the continuous paramagnetic-to-antiferromagnetic transition.
(f)~Same as (c), but along Path~4, as functions of $r_\phi$ with $r_T=r_\phi$, across the first-order paramagnetic-to-antiferromagnetic transition.
}
\label{fig:mean-field-paths}
\end{figure}
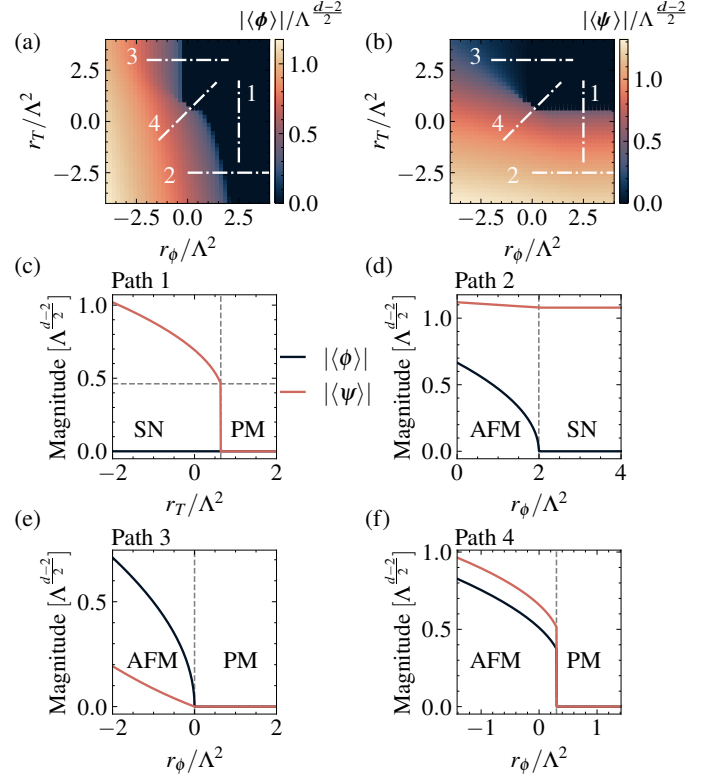

We first determine the mean-field phase diagram of the model defined in Eq.~\eqref{eq:full-Hamiltonian}, neglecting fluctuations of both the vector and tensor fields. To this end, we consider the corresponding Landau free-energy functional and numerically minimize it with respect to $\boldsymbol{\phi}$ and $\boldsymbol{\psi}$.

\subsection{Phase diagram}

As an example, we fix the dimensionless couplings $(u_\phi\Lambda^{d-4},v_\phi\Lambda^{d-4},u_T\Lambda^{d-4},\omega\Lambda^{(d-6)/2},\lambda\Lambda^{(d-6)/2})=(24,12,36,7.2,1.6)$.
The resulting mean-field order parameters $|\langle\boldsymbol{\phi}\rangle|$ and $|\langle\boldsymbol{\psi}\rangle|$ as functions of $r_\phi$ and $r_T$ are shown in Figs.~\ref{fig:mean-field-PD}(a) and \ref{fig:mean-field-PD}(b), respectively. 
Similar phase diagrams are obtained for other choices of the couplings. The mean-field theory yields three phases: a symmetric paramagnetic phase (PM) with $\langle\boldsymbol{\psi}\rangle=\langle\boldsymbol{\phi}\rangle=0$, a spin-nematic phase (SN) with $\langle\boldsymbol{\psi}\rangle\neq0$ and $\langle\boldsymbol{\phi}\rangle=0$, and an antiferromagnetic phase (AFM) with $\langle\boldsymbol{\psi}\rangle\neq0$ and $\langle\boldsymbol{\phi}\rangle\neq0$. All three phases meet at the triple point A [see Fig.~\ref{fig:mean-field-PD}(a,b)], whose coordinates can be obtained analytically,
\begin{equation}
\label{eq:point-A}
\mathrm A: \quad (r_\phi,r_T) \Bigr|_A = \left(\frac{8\omega\lambda}{3u_T}, \frac{4\omega^2}{9u_T}\right).
\end{equation}

\subsection{Phase transitions}

The order of the transitions can be determined by following specific paths through the mean-field phase diagram, as indicated in Figs.~\ref{fig:mean-field-paths}(a) and \ref{fig:mean-field-paths}(b). 

\paragraph{Paramagnetic-to-nematic transition.}

In Fig.~\ref{fig:mean-field-paths}(c), we show the magnitudes of the two order parameters along Path~1, tuning the nematic mass $r_T$ across the paramagnetic-to-nematic phase boundary at fixed $r_\phi/\Lambda^2=2.5$. Along this path, the antiferromagnetic order parameter (black) remains zero, while the nematic order parameter (orange) jumps to a finite value at $r_{T}=r_T\bigr|_\mathrm{A}$, revealing a first-order transition. The same behavior is obtained for other values of $r_\phi$. In fact, the critical value of $r_T$ is independent of $r_\phi$, so that the phase boundary is exactly horizontal and is given by 
\begin{align}
\text{PM-to-SN:}
\quad r_\phi>r_\phi\bigr|_\mathrm{A}, \quad r_T=r_T\bigr|_\mathrm{A}\,,
\end{align}
as indicated by the horizontal dashed line in Fig.~\ref{fig:mean-field-PD}(b). The magnitude of the jump, $\lim_{r_T \to r_T |_\mathrm{A}^-}|\langle\boldsymbol{\psi}\rangle|=\frac{4\omega}{\sqrt{3}u_T}$, is likewise independent of $r_\phi$. The first-order nature of the paramagnetic-to-nematic transition in dimensions above two is consistent with the equivalence of the nematic Hamiltonian [Eq.\eqref{eq:nematic-part}] to the four-state Potts model~\cite{zia75,wu82}.

\paragraph{Nematic-to-antiferromagnetic transition.}

The transition from the nematic to the antiferromagnetic phase is illustrated along Path~2 in Fig.~\ref{fig:mean-field-paths}(d), where we fix the tensor mass to $r_T/\Lambda^2=-2.5$ and tune the vector mass $r_\phi$. The transition is continuous, with the antiferromagnetic order parameter growing continuously from zero as a power law. The nematic order parameter, which is already finite on both sides of the transition, remains continuous but changes slope at the transition, as shown in Fig.~\ref{fig:mean-field-paths}(d). The corresponding phase boundary is given by
\begin{equation}
\text{SN-to-AFM:}
\quad
r_\phi > r_\phi \bigr|_\mathrm{A}\,,
r_T = -\frac{u_T}{2\lambda^2}\left(\frac{r_\phi}{2}-\frac{\omega\lambda}{u_T}\right)^2 
+ \frac{\omega^2}{2u_T}\,,
\end{equation}
and forms the parabolic continuous phase boundary shown in Fig.~\ref{fig:mean-field-PD}(a). The continuous nature of the transition is consistent with the fact that, in a nematic background, the remaining fluctuating vector degrees of freedom are effectively described by a $\mathbb{Z}_2$-symmetric Ising model; see Appendix~\ref{app:rt-minus-infty}. The transition is therefore expected to belong to the Ising universality class.

\paragraph{Paramagnetic-to-antiferromagnetic transition.}

The nature of the direct paramagnetic-to-antiferromagnetic transition depends on the value of $r_T$ at the transition. This is illustrated in Figs.~\ref{fig:mean-field-paths}(e) and \ref{fig:mean-field-paths}(f), which show the magnitudes of the order parameters along Paths~3 and 4, respectively.
Along Path~3, we tune $r_\phi$ at fixed, sufficiently large $r_T/\Lambda^2=3$. The transition is direct and continuous, as shown in Fig.~\ref{fig:mean-field-paths}(e). The corresponding phase boundary is exactly vertical and given by
\begin{equation}
\label{eq:symmetric-AFM-boundary}
\text{PM-to-AFM (continuous part):}
\quad
r_\phi=0\,,
\quad 
r_T>r_T \bigr|_\mathrm{B}\,,
\end{equation}
as indicated by the vertical continuous line in Fig.~\ref{fig:mean-field-PD}(a). Here, B denotes the point at which the direct PM-to-AFM transition changes from continuous to first order. Its coordinates are
\begin{equation}
\mathrm{B}: \quad (r_\phi,r_T) \Bigr|_\mathrm{B} =\left(0,\frac{4\omega^2}{9u_T}+\frac{12\lambda^2}{3u_\phi+v_\phi}\right)\,.
\end{equation}
An example of a direct first-order transition is shown in Fig.~\ref{fig:mean-field-paths}(f), where we tune along Path~4, defined by $r_T=r_\phi$, cf.~Figs.~\ref{fig:mean-field-paths}(a) and \ref{fig:mean-field-paths}(b). Both the nematic and antiferromagnetic order parameters jump at the transition, demonstrating its discontinuous nature.
The first-order part of the paramagnetic-to-antiferromagnetic phase boundary extends from A to B,
\begin{equation}
\text{PM-to-AFM (first-order part):}
\quad
\mathcal C_{\text{A} \to \text{B}}.
\end{equation}
Its analytical parametrization is more involved, as it requires solving a cubic equation, and does not provide further insight. We therefore determine this boundary numerically; it is shown as a dashed line connecting A and B in Figs.~\ref{fig:mean-field-PD}(a) and \ref{fig:mean-field-PD}(b).
Across the direct paramagnetic-to-antiferromagnetic transition, the $S_4$ and $\mathbb Z_2$ subgroups of the cubic symmetry group ${O}_h\simeq S_4\times\mathbb Z_2$ are broken simultaneously. Once the primary order parameter $\langle\boldsymbol{\phi}\rangle$ condenses, its bilinears act as a conjugate field for $\boldsymbol{\psi}$ and therefore generically induce a finite secondary order parameter.
In fact, throughout the antiferromagnetic phase, including on its phase boundaries, the two order parameters satisfy the constraint
\begin{equation}
    \label{eq:fully-order-OP-constraint}
    \frac{2\lambda}{\sqrt{3}}|\langle\boldsymbol{\psi}\rangle|=r_\phi+\frac{3u_\phi+v_\phi}{18}|\langle\boldsymbol{\phi}\rangle|^2\,,
\end{equation}
where the magnitudes $|\langle\boldsymbol{\psi}\rangle|$ and $|\langle\boldsymbol{\phi}\rangle|$ implicitly depend on the system parameters. At the continuous paramagnetic-to-antiferromagnetic transition, the primary order parameter vanishes as $|\langle\boldsymbol{\phi}\rangle|\propto |r_\phi|^{1/2}$. Equation~\eqref{eq:fully-order-OP-constraint} then implies $|\langle\boldsymbol{\psi}\rangle|\propto|\langle\boldsymbol{\phi}\rangle|^2\propto |r_\phi|$ close to the transition. Thus, the secondary nematic order parameter has the mean-field exponent $\beta_\psi=1$, twice the exponent $\beta_\phi=1/2$ of the primary order parameter, as expected for an order parameter induced by a bilinear of the primary field.

The case $\omega=0$ might naively suggest a continuous paramagnetic-to-nematic transition at the mean-field level. However, $\omega=0$ represents a fine-tuned situation that can occur only in the presence of additional symmetries. Moreover, setting $\omega=0$ does not guarantee a continuous transition. In all cases we have studied, the order parameters exhibit finite jumps, similar to those observed along Path~4 in Fig.~\ref{fig:mean-field-paths}(f), indicating that the transition along the phase boundary connecting A and B in Figs.~\ref{fig:mean-field-PD}(a) and \ref{fig:mean-field-PD}(b) remains first order even for $\omega=0$.

\paragraph{Discussion.}

Let us summarize the nature of the phase transitions. The paramagnetic-to-nematic transition is generically first order, whereas the nematic-to-antiferromagnetic transition is continuous. The paramagnetic and antiferromagnetic phases are separated by a continuous transition above point B and a first-order transition below it. First-order and continuous transitions are indicated by dashed and solid lines, respectively, in Figs.~\ref{fig:mean-field-PD}(a) and \ref{fig:mean-field-PD}(b).

The neglect of fluctuations inherent in the mean-field approximation is justified only above the upper critical dimension of the model defined in Eq.~\eqref{eq:full-Hamiltonian}. To identify this dimension, we list the engineering scaling dimensions of the fields and couplings in Table~\ref{tab:tree-level-scaling}.
\begin{table}[t]
\caption{Engineering scaling dimensions of the fields and couplings in the Hamiltonian in Eq.~\eqref{eq:full-Hamiltonian}, expressed in units of inverse length.}
\centering
\begin{tabular*}{\linewidth}{@{\extracolsep{\fill}}cc}
\hline
\hline
Field or coupling & Engineering scaling dimension \\
\hline
$\boldsymbol{\phi}$, $\boldsymbol{\psi}$ & $\frac{d-2}{2}$ \\
$r_\phi$, $r_T$ & $2$ \\
$u_\phi$, $v_\phi$, $u_T$ & $4-d$ \\
$\omega$, $\lambda$ & $\frac{6-d}{2}$ \\[3pt]
\hline
\hline
\end{tabular*}
\label{tab:tree-level-scaling}
\end{table}
In particular, the couplings $\omega$ and $\lambda$, which characterize the four-state Potts interaction in the nematic sector and the vector-nematic interaction, respectively, become irrelevant above six dimensions. The upper critical dimension is therefore $d_\mathrm{c}=6$. Fluctuations must consequently be taken into account beyond mean-field theory in the physical dimensions $d=2$ and $d=3$. To this end, we employ a RG approach based on an $\epsilon$ expansion about the upper critical dimension, complemented by a perturbative analysis in fixed dimensions.

\section{Renormalization group analysis}
\label{sec:renormalization-group}

We employ a Wilsonian RG approach at one-loop order in fixed spatial dimension $d$~\cite{herbut07}. Importantly, fluctuations of the nematic and antiferromagnetic order parameters are treated on equal footing. We introduce dimensionless couplings through the rescaling
\begin{align}
(r_\phi, r_T) \Lambda^{-2} & \mapsto (r_\phi, r_T)\,,\displaybreak[0] \label{eq:rescaling-1} \\
(u_\phi,v_\phi,u_T)\frac{S_d \Lambda^{d-4}}{(2\pi)^d} &\mapsto (u_\phi,v_\phi,u_T) \,,\displaybreak[0] \\
(\omega^2,\lambda^2)\frac{S_d \Lambda^{d-6}}{(2\pi)^d} &\mapsto (\omega^2,\lambda^2) \,, \label{eq:rescaling-3}
\end{align}
consistent with the engineering scaling dimensions listed in Table~\ref{tab:tree-level-scaling}. Here, $\Lambda$ denotes the ultraviolet momentum cutoff, which may be identified with the inverse of the shortest microscopic length scale, such as the lattice spacing $a$, so that $\Lambda\sim a^{-1}$. We then simultaneously integrate out the fast modes of the vector and tensor fields with momenta in the shell $\Lambda/b\leq|\mathbf{k}|\leq\Lambda$, where $b>1$, yielding the corresponding RG flow equations in terms of the $\eta$ and $\beta$ functions.

\subsection{Flow equations}

\paragraph{Eta functions.}

\begin{figure}[b!]
\centering
\begin{overpic}[width=\columnwidth]{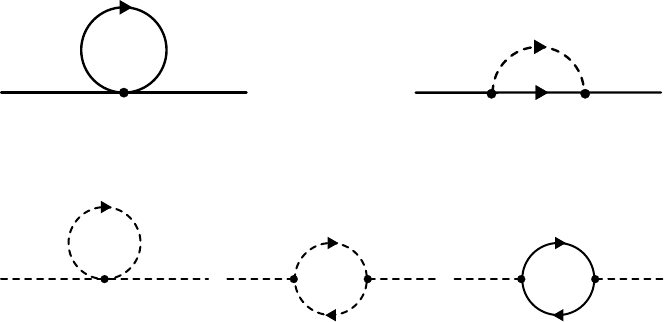}
\put(0,48){(a)}
\put(62,48){(b)}
\put(14,31){$u_\phi$, $v_\phi$}
\put(71,31){$\lambda$}
\put(89,31){$\lambda$}
\put(0,18){(c)}
\put(34,18){(d)}
\put(68,18){(e)}
\put(14,3){$u_T$}
\put(41,3){$\omega$}
\put(56,3){$\omega$}
\put(76,3){$\lambda$}
\put(90,3){$\lambda$}
\end{overpic}
\caption{%
One-loop diagrams contributing to the vector (top row) and tensor (bottom row) two-point functions. Solid and dashed lines represent the vector and tensor fields, respectively.
The tadpole diagrams in (a) and (c) contribute to $\beta_{r_\phi}$ and $\beta_{r_T}$, respectively, but not to the anomalous dimensions $\eta_\phi$ and $\eta_T$.
}
\label{fig:two-point-diagrams}
\end{figure}
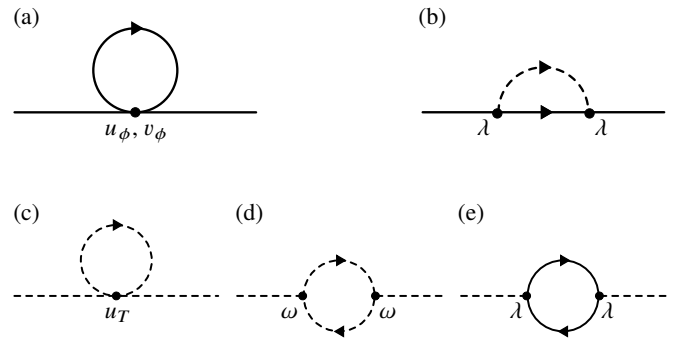

The $\eta$ functions describe the corrections to the coefficients of the $\mathbf{k}^2$ terms in the propagators and are obtained at one-loop order by evaluating the diagrams shown in Fig.~\ref{fig:two-point-diagrams}. We find, using $N_\phi = N_T = 3$,
\begin{widetext}
\begin{align}
    \label{eq:eta-functions-1}
        \eta_\phi &=-\frac{2\lambda^2}{d}\left(\frac{(1-a^2)(1+r_\phi)^2(4-d-dr_T) + a^2(1+r_T)^2(4-d-dr_\phi)-4a(1-a)(1+r_T)(1+r_\phi)}{(1+r_T)^3(1+r_\phi)^3} \right)\,, \\
        \eta_T &= -\frac{\omega^2}{d}\left(\frac{(4-d-dr_T)[(1-a)^2+a^2]-4a(1-a)}{(1+r_T)^4} \right) -\frac{\lambda^2}{d}\left(\frac{(4-d-dr_\phi)[(1-a)^2+a^2]-4a(1-a)}{(1+r_\phi)^4}\right)\,,
\end{align}
\end{widetext}
Here, $a\in[0,1]$ parametrizes the distribution of the external momentum among the internal lines in the loop diagrams shown in Figs.~\ref{fig:two-point-diagrams}(b), \ref{fig:two-point-diagrams}(d), and \ref{fig:two-point-diagrams}(e). The choice $a=1/2$ corresponds to a symmetric distribution~\cite{janssen15}, while $a=0$ and $a=1$ give maximally asymmetric distributions.
We have verified that the resulting flow equations become independent of $a$ in the vicinity of the upper critical dimension, but acquire a weak $a$ dependence in fixed dimensions away from $d=6$.
For our fixed-dimensional numerical results, we set $a=1/2$.
At a fixed point, the $\eta$ functions evaluated at the fixed-point couplings give the corresponding anomalous dimensions $\eta_\phi$ and $\eta_T$ of the vector and tensor fields. We note that the tadpole diagrams in Figs.~\ref{fig:two-point-diagrams}(a) and \ref{fig:two-point-diagrams}(c) do not contribute to the anomalous dimensions, as in conventional $\phi^4$ theory. In the present model, however, the cubic $\omega$ and vector-tensor $\lambda$ couplings generate nontrivial one-loop corrections to the anomalous dimensions through the diagrams in Figs.~\ref{fig:two-point-diagrams}(b), \ref{fig:two-point-diagrams}(d), and \ref{fig:two-point-diagrams}(e), similar to other theories with cubic interactions~\cite{fei14,herbut16}.

\paragraph{Beta functions.}

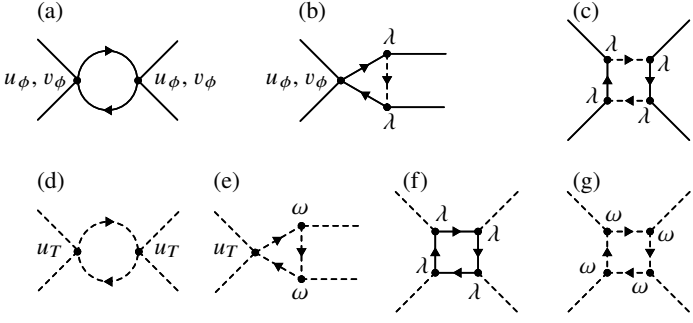
\begin{figure}[tb]
\centering
\begin{overpic}[width=\columnwidth]{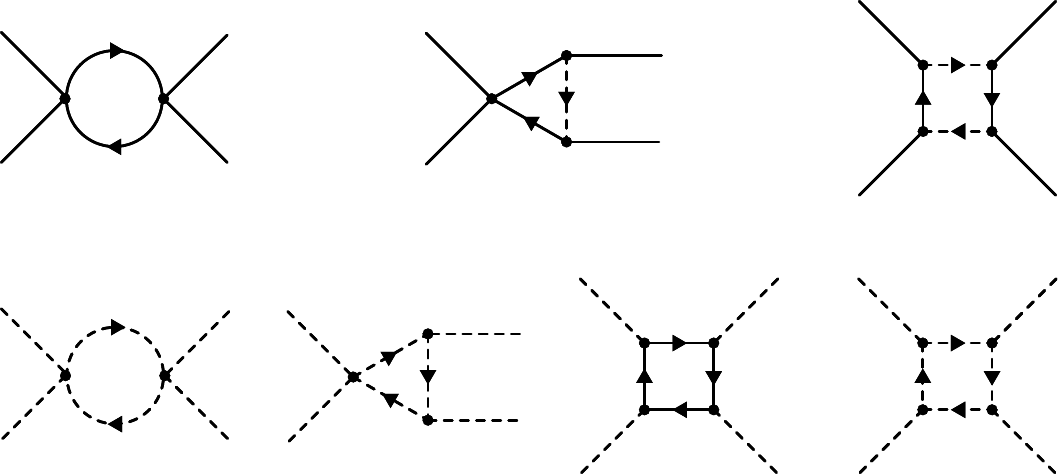}
\put(0,45){(a)}
\put(40,45){(b)}
\put(82,45){(c)}
\put(0,19){(d)}
\put(27,19){(e)}
\put(56,19){(f)}
\put(82,19){(g)}
\put(-5,35){$u_\phi$, $v_\phi$}
\put(18,35){$u_\phi$, $v_\phi$}
\put(35,35){$u_\phi$, $v_\phi$}
\put(53,41){$\lambda$}
\put(53,28){$\lambda$}
\put(84,32){$\lambda$}
\put(87,40){$\lambda$}
\put(95,37){$\lambda$}
\put(92,29){$\lambda$}
\put(0,9){$u_T$}
\put(18,9){$u_T$}
\put(27,9){$u_T$}
\put(39,15){$\omega$}
\put(39,2){$\omega$}
\put(58,6){$\lambda$}
\put(61,14){$\lambda$}
\put(69,11){$\lambda$}
\put(66,2){$\lambda$}
\put(83,6){$\omega$}
\put(87,14){$\omega$}
\put(95,11){$\omega$}
\put(91,3){$\omega$}
\end{overpic}
\caption{%
One-loop diagrams contributing to the flow of the four-point couplings $u_\phi$ and $v_\phi$ (top row) and $u_T$ (bottom row). Solid and dashed lines represent the vector and tensor fields, respectively.
The diagrams in (a) and (b) contribute to both $\beta_{u_\phi}$ and $\beta_{v_\phi}$, while the diagram in (c) contributes only to $\beta_{u_\phi}$.
}
\label{fig:four-point-diagrams}
\end{figure}

\begin{figure}[tb!]
\centering
\begin{overpic}[width=\columnwidth]{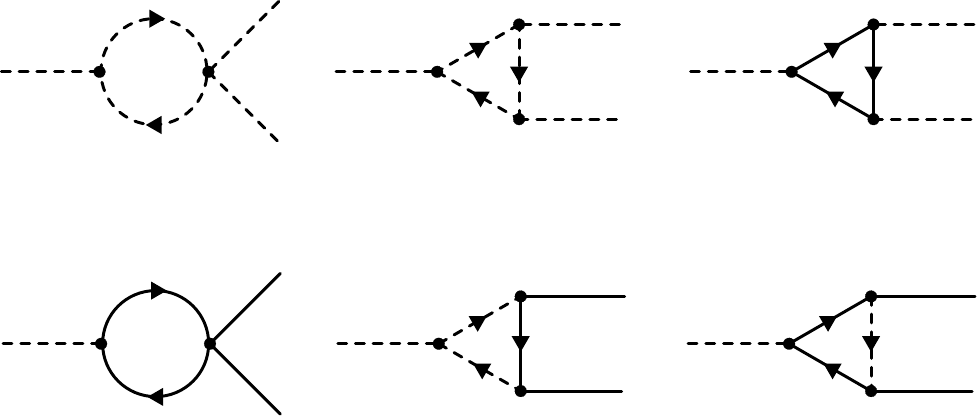}
\put(0,45){(a)}
\put(33,45){(b)}
\put(70,45){(c)}
\put(0,18){(d)}
\put(33,18){(e)}
\put(70,18){(f)}
\put(7,36){$\omega$}
\put(23,34.5){$u_T$}
\put(42,36){$\omega$}
\put(52,42){$\omega$}
\put(52,27){$\omega$}
\put(78,36){$\lambda$}
\put(88,42){$\lambda$}
\put(88,26){$\lambda$}
\put(7,8){$\lambda$}
\put(23,6.5){$u_\phi$, $v_\phi$}
\put(42,8){$\omega$}
\put(52,14){$\lambda$}
\put(52,-1){$\lambda$}
\put(78,8){$\lambda$}
\put(88,14){$\lambda$}
\put(88,-1){$\lambda$}
\end{overpic}
\caption{%
One-loop diagrams contributing to the flow of the three-point couplings $\omega$ (top row) and $\lambda$ (bottom row).
Solid and dashed lines represent the vector and tensor fields, respectively.
}
\label{fig:three-point-diagrams}
\end{figure}
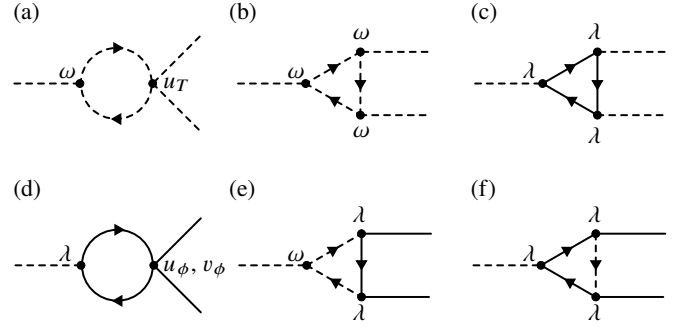

The $\beta$ functions are obtained by evaluating the one-loop diagrams shown in Figs.~\ref{fig:two-point-diagrams}--\ref{fig:three-point-diagrams} at vanishing external momentum. For a generic parameter $g\in{r_\phi,r_T,u_\phi,v_\phi,u_T,\omega,\lambda}$, the corresponding $\beta$ function takes the form
\begin{equation}
    \label{eq:generic-beta-function}
    \beta_{g}(r_\phi,r_T,u_\phi,v_\phi,u_T,\omega,\lambda)=\frac{\dd g}{\dd \ln{b}}=\left(\left[g\right]-\eta_g\right)g+\dots
\end{equation}
where $[g]$ denotes the engineering scaling dimension of $g$ listed in Table~\ref{tab:tree-level-scaling}, and $\eta_g$ is the corresponding $\eta$ function, which accounts for the anomalous scaling generated by fluctuations. The ellipsis denotes the explicit vertex-renormalization contributions.

Evaluating the diagrams in Fig.~\ref{fig:two-point-diagrams} yields the flow equations for the masses of the vector and tensor fields, $r_\phi$ and $r_T$,
\begin{align}
        \beta_{r_\phi}&=(2-\eta_\phi)r_\phi+\frac{(N_\phi+2)u_\phi+v_\phi}{6(1+r_\phi)}-\frac{2\lambda^2}{(1+r_\phi)(1+r_T)}\label{eq:rPhi-beta-func}\,, \displaybreak[0] \\
        \beta_{r_T}&=(2-\eta_T)r_T+\frac{(N_T+2)u_T}{6(1+r_T)}-\frac{\omega^2}{(1+r_T)^2}-\frac{\lambda^2}{(1+r_\phi)^2}\label{eq:rT-beta-func}\,.
\end{align}
The first two terms in each equation have the usual form of the mass flow in an $\mathrm{O}(N)$ model with $N=N_\phi$ and $N = N_T$, respectively.
In the vector sector, the cubic anisotropy $v_\phi$ provides an additional contribution to the tadpole correction. The remaining terms arise from the three-point interactions: the vector-tensor coupling $\lambda$ contributes to both mass flows, while the cubic tensor coupling $\omega$ contributes only to $\beta_{r_T}$.
To facilitate comparison with results in the literature, we have explicitly retained the factors of $N_\phi$ and $N_T$ in the terms involving $u_\phi$ and $u_T$ in the equations above. We emphasize, however, that the vector-tensor coupling is hard-wired to $N_\phi=N_T=3$, and hence so are all terms in the flow equations involving $\lambda$.

Evaluating the diagrams in Fig.~\ref{fig:four-point-diagrams} yields the flow equations for the four-point couplings $u_\phi$, $v_\phi$, and $u_T$,
\begin{align}
\beta_{u_\phi}&=(4-d-2\eta_\phi)u_\phi-\frac{(N_\phi+8)u_\phi^2+6u_\phi v_\phi}{6(1+r_\phi)^2}
\nonumber\\&\quad
+ \frac{2\lambda^2[(N_\phi+5)u_\phi+3v_\phi]}{(1+r_\phi)^2(1+r_T)} -\frac{4\lambda^4(N_\phi+3)}{(1+r_\phi)^2(1+r_T)^2}\,, 
\label{eq:uPhi-beta-func}\displaybreak[0]\\
\beta_{v_\phi}&=(4-d-2\eta_\phi)v_\phi -\frac{3v_\phi^2+4u_\phi v_\phi}{2(1+r_\phi)^2}
\nonumber\\&\quad
+\frac{6v_\phi\lambda^2}{(1+r_\phi)^2(1+r_T)}\,,
\label{eq:vPhi-beta-func}\displaybreak[0]\\
\beta_{u_T}&=(4-d-2\eta_T)u_T-\frac{(N_T+8)u_T^2}{6(1+r_T)^2}+\frac{12u_T\omega^2}{(1+r_T)^3}
\nonumber\\&\quad 
-\frac{3(N_T+3)\omega^4}{(1+r_T)^4}-\frac{3(N_\phi+3)\lambda^4}{(1+r_\phi)^4}\,.
\label{eq:uT-beta-func}
\end{align}
The terms quadratic in the four-point couplings reproduce the familiar $\mathrm O(N_\phi)$ and $\mathrm O(N_T)$ contributions, with the cubic anisotropy $v_\phi$ providing an additional contribution in the vector sector. The $\omega$ and $\lambda$ couplings generate further contributions through the three-point interactions, including terms of order $\omega^2$, $\omega^4$, $\lambda^2$, and $\lambda^4$. The latter two are, again, hard-wired to $N_\phi=N_T=3$.

Evaluating the diagrams in Fig.~\ref{fig:three-point-diagrams} yields the flow equation for the three-point couplings $\omega$ and $\lambda$ for $N_\phi=N_T=3$,
\begin{align}
\beta_{\omega}&=\left(\frac{6-d-3\eta_T}{2}\right)\omega -\frac{u_T\omega}{(1+r_T)^2}+\frac{\omega^3}{(1+r_T)^3}+\frac{\lambda^3}{(1+r_\phi)^3}\label{eq:omega-beta-func}\,,\displaybreak[0]\\
\beta_\lambda &=\left[\frac{6-d-(2\eta_\phi+\eta_T)}{2} \right]\lambda -\frac{\lambda u_\phi}{3(1+r_\phi)^2}
\nonumber\\&\quad
+\frac{\omega \lambda^2}{(1+r_\phi)(1+r_T)^2}+\frac{\lambda^3}{(1+r_T)(1+r_\phi)^2}\,.
\label{eq:lambda-beta-func}
\end{align}
In particular, $\beta_\omega$ contains a term proportional to $\lambda^3$. Thus, even if the four-state anisotropy is absent in the bare nematic sector, corresponding to the fine-tuned case $\omega=0$, fluctuations generate a nonzero $\omega$ through the vector-nematic interaction. This occurs because no symmetry protects the condition $\omega=0$ under the RG flow. For $\lambda>0$, the generated $\omega$ is positive, favoring the nematic configurations illustrated in Fig.~\ref{fig:minima-sketches}(d).

The zeros of the $\beta$ functions correspond to the fixed points of the RG flow. While closed-form expressions for the fixed points cannot be obtained in general dimension $d$, we analyze their structure using several complementary approaches. First, we develop a controlled $\epsilon$ expansion about the upper critical dimension $d_{\mathrm{c}}=6$. We then numerically track the fixed points to lower dimensions. Finally, we consider the limits $r_T\to\infty$, $r_\phi\to\infty$, and $r_T\to-\infty$, in which the model in Eq.~\eqref{eq:full-Hamiltonian} reduces to three simpler theories that are amenable to nonperturbative analysis in low dimensions, in particular in $d=2$.

\subsection{Fixed-point analysis}
\label{subsec:eps-expansion}

\begin{table*}[bt!]
\caption{%
Perturbative fixed points (FPs) with $|r_\phi|,|r_T|<\infty$ that may describe the phase transition across the triple point A, together with their locations and number of RG relevant directions (``\# rel.~dir.'') in $d=6-\epsilon$ dimensions for $0<\epsilon\ll1$. The fixed point with only two relevant directions occurs at imaginary $\lambda_\star$ and $\omega_\star$, suggesting a first-order transition across the triple point A and the absence of critical fluctuations about the metastable state for $d<6$.
}
\centering
\begin{tabular*}{\linewidth}{@{\extracolsep{\fill}}l l p{5em} p{14em}}
\hline
\hline
FP & $(r_{\phi\star}, r_{T\star}, u_{\phi\star}, v_{\phi\star}, u_{T\star}, \omega^2_{\star}, \lambda^2_{\star})$ & \# rel.~dir.\newline ($d=6-\epsilon$) & Comments\\
\hline
I & $(0, 0, 0, 0, 0, 0, 0)$ & 4 & Gaussian fixed point, multicritical for $d>6$\\
II & $\left(-3\epsilon, -\frac{3}{2}\epsilon, -18(N_\phi+3)\epsilon^2, 0, -\frac{27}{2}(N_\phi+3)\epsilon^2, 0, -3\epsilon\right)$ & 3\\
III & $\left(-\frac{\epsilon}{2}, -\frac{\epsilon}{2}, -\frac{1}{2}(N_\phi+3)\epsilon^2, 0, -\frac{3}{8}(N_\phi+N_T+6)\epsilon^2, -\frac{\epsilon}{2}, -\frac{\epsilon}{2}\right)$ & 3 \\
IV & $\left(-\frac{12}{31}\epsilon, -\frac{39}{62}\epsilon, -\frac{288}{961}(N_\phi+3)\epsilon^2, 0, -\frac{27}{1922}(16N_\phi+81N_T+291)\epsilon^2, -\frac{27}{31}\epsilon, -\frac{12}{31}\epsilon\right)$ & 2 & Complex interacting fixed point, multicritical for $d<6$\\
V & $\left(0, -\frac{\epsilon}{2}, 0, 0, -\frac{3}{2}(N_T+3)\epsilon^2, -\epsilon, 0\right)$ & 3 & Decoupled Potts fixed point\\[3pt]
\hline
\hline
\end{tabular*}
\label{tab:epsilon-FP}
\end{table*}

\subsubsection{Triple point (\texorpdfstring{$|r_\phi|, |r_T| < \infty$}{|rphi|, |rT| < infty})}
\label{subsubsec:triple-point}

We begin by analyzing the possibility of multicritical behavior at the triple point A, where the antiferromagnetic, nematic, and paramagnetic phases meet [see Fig.~\ref{fig:mean-field-PD}(a,b)]. To this end, we search for fixed-point solutions with two RG-relevant directions, corresponding to the two tuning parameters required to reach the triple point. 

\paragraph{Above six dimensions.}

For $d>d_\mathrm{c}$, we expect mean-field theory to become asymptotically exact in the critical region. Indeed, in this regime, the Gaussian fixed point at $(r_{\phi\star},r_{T\star},u_{\phi\star},v_{\phi\star},u_{T\star},\omega^2_\star,\lambda^2_\star)=(0,0,0,0,0,0,0)$ has only two RG-relevant directions, corresponding to $r_\phi$ and $r_T$, and can therefore describe a possible multicritical point.
We emphasize, however, that the existence of a Gaussian fixed point with two relevant directions does not by itself guarantee a continuous transition in the mean-field phase diagram. First, the system may lie outside its basin of attraction and consequently exhibit a first-order transition. Second, the state described by the Gaussian fixed point may be metastable, in which case the RG flow in its vicinity describes critical fluctuations around the metastable state rather than the true thermodynamic state~\cite{priest76}. We argue that the latter scenario is realized here, with the metastability arising from the cubic term $\propto\psi_a\psi_b\psi_c$ in the tensor Hamiltonian, Eq.~\eqref{eq:nematic-part}. In Appendix~\ref{app:metastable}, we illustrate, using a simple $\phi^3$ field theory, how critical fluctuations about a metastable state are described by the RG flow in the vicinity of a Gaussian fixed point.

\paragraph{Below six dimensions.}

For $d<d_\mathrm{c}$, the Gaussian fixed point develops two additional RG-relevant directions, associated with the cubic couplings $\lambda$ and $\omega$. To obtain analytical control over the fixed-point equations in the perturbative domain, we introduce the small parameter $\epsilon=6-d$ and first consider the regime of small $\epsilon$. We then track the resulting fixed-point solutions to lower dimensions numerically.
Specifically, we expand each of the seven parameters, $g\in\{r_\phi,r_T,u_\phi,v_\phi,u_T,\omega^{2},\lambda^{2}\}$, at the fixed point as $g_\star=\sum_{n=0}^\infty g^{(n)}_\star\epsilon^n$ and substitute these expansions into the fixed-point equations. Since the quartic couplings $u_\phi$, $v_\phi$, and $u_T$ are irrelevant above four dimensions, we expect their fixed-point values to arise only at higher order in $\epsilon$, whereas the squares of the cubic couplings, $\omega^2$ and $\lambda^2$, are expected to be of linear order in $\epsilon$ at nontrivial fixed points~\cite{priest76, janssen15}. Truncating the resulting equations at the first nontrivial order in $\epsilon$ allows them to be solved analytically, yielding the perturbative fixed-point values listed in Table~\ref{tab:epsilon-FP}.

Except for the Gaussian fixed point, labeled ``I'' in Table~\ref{tab:epsilon-FP}, all fixed points have negative values of $\lambda_\star^2$ and $\omega_\star^2$, corresponding to purely imaginary fixed-point values of $\lambda_\star$ and $\omega_\star$. This includes the potential multicritical fixed point with two relevant directions, labeled ``IV'' in Table~\ref{tab:epsilon-FP}. 
The conformal field theories (CFTs) defined by these fixed points are examples of ``imaginary CFTs'' in the sense of Ref.~\cite{wiese24}, representing critical theories after rotating both fields independently as $\phi \mapsto i \phi$ and $\psi \mapsto i \psi$.
Since the RG flow starts from real values of the couplings and the flow equations have real coefficients, these imaginary fixed points cannot be reached from the physical parameter space. This implies the absence of any critical fluctuations about any stable or metastable state, similar to the situation in the $\phi^3$ field theory below the upper critical dimension, cf.~Appendix~\ref{app:metastable} and Refs.~\cite{priest-76-percolation,fisher78-phi3,gracey15-phi3,borinsky21-phi3}.
At the fixed point labeled ``V'' in Table~\ref{tab:epsilon-FP}, the vector and tensor sectors decouple, $\lambda_\star=0$, and only the tensor couplings acquire nontrivial fixed-point values. This fixed point can therefore be viewed as the $r_{\phi\star}=0$ counterpart of the Potts fixed point, which we discuss further in Sec.~\ref{subsubsec:nem-to-PM}.

Our flow equations are formulated in arbitrary fixed dimension $d$, albeit within a low-order perturbative expansion. In principle, this allows us to determine the fixed points numerically for arbitrary $d$. From this perspective, the $\epsilon$ expansion provides a controlled approach near the upper critical dimension $d_\mathrm{c}=6$. For $4<d<6$, however, our perturbative treatment neglects higher-order diagrams generated by fluctuations of $\omega$ and $\lambda$, which may become relevant away from $d_\mathrm{c}$. Moreover, $d=4$ constitutes a second critical dimension, below which the quartic couplings $u_\phi$, $v_\phi$, and $u_T$ become relevant already at tree level. We may therefore expect the RG behavior to change qualitatively also upon crossing $d=4$.
To gain insight into the physics away from the upper critical dimension, we track the fixed-point solutions found in the $6-\epsilon$ expansion numerically to dimensions below $d=4$. In all dimensions investigated, we find no real physical multicritical fixed point: the fixed-point values of $\lambda_\star^2$ and $\omega_\star^2$ remain negative throughout.

From these results, we conclude that critical fluctuations about the metastable state described by the Gaussian fixed point exist for $d\geq6$, but no corresponding critical fluctuations occur in the physical regime $d<6$.
We emphasize that this conclusion is controlled only sufficiently close to the upper critical dimension. Nevertheless, as discussed below, comparison with nonperturbative results in $d=2$ indicates that the fixed-dimension perturbative RG analysis employed here captures the qualitative behavior also at lower dimensions. We therefore expect the thermodynamic transition at the triple point A in the phase diagram [Fig.~\ref{fig:mean-field-PD}(a,b)] to remain generically first order in all dimensions $d \geq 2$.

\subsubsection{Paramagnetic-to-antiferromagnetic transition (\texorpdfstring{$r_T\rightarrow\infty$}{rT->infty})}
\label{subsubsec:AFM-to-PM}

\begin{table*}[tb!]
\caption{%
Selected fixed points (FPs) for $r_{T} \to \infty$ that may describe the antiferromagnetic-to-paramagnetic transition, with their locations and number of RG-relevant directions (``\# rel.~dir.'') near and below four dimensions.
Here, $N \equiv N_\phi$ denotes the number of components of the vector field $\phi$ and $f \equiv f(d,a)$, $g \equiv g(N_\phi,d,a)$, $h \equiv h(N_\phi,d)$, $j \equiv j(N_\phi,d,a)$, $k \equiv k(N_\phi,d)$, $m \equiv m(N_\phi,d)$, $p \equiv p(N_\phi,d)$, $t \equiv t(d,a)$, and $w \equiv w(N_\phi,d)$ are functions of the dimension $d$, the number of vector components $N_\phi$, and the momentum-distribution parameter $a$. Their explicit forms are given in Appendix~\ref{app:placeholders}.
For $d<4$, the critical fixed point depends on $N \equiv N_\phi$: For $N < N_\mathrm{c}$ ($N > N_\mathrm{c}$), the cubic anisotropy is irrelevant (relevant) at criticality, and the transition is governed by the O($N_\phi$) fixed point (cubic fixed point). At one-loop order, we find $N_\mathrm{c} = 4$, but higher-loop corrections reduce $N_\mathrm{c}$ for $d=3$ to a value slightly below 3. For $d=2$, we have $N_\mathrm{c} = 2$.
}
\centering
\begin{tabular*}{\linewidth}{@{\extracolsep{\fill}}l l p{5.5em} p{12em}}
\hline
\hline
FP& $(r_{\phi\star}, r_{T\star}, u_{\phi\star}, v_{\phi\star}, u_{T\star}, \omega^2_{\star}, \lambda^2_{\star})$ & \# rel.~dir.\newline ($d = 4 - \epsilon$) & Comments\\
\hline
VI & $\left(0, \infty, 0, 0, 0, 0, 0\right)$ & 6 & Gaussian fixed point, critical for $d>6$\\
VII & $\left(0, \infty, 0, 0, \frac{3(N+3)(6-d)^2 f^2}{d-8}, (6-d)f^3, (6-d)f\right)$ & 3 & Gaussian$'$ fixed point, critical for $4<d<6$\\
VIII&$\left(\frac{(4-d) (N+2)}{(d-6) N+2 (d-12)},
\infty,
\frac{24(4-d)(N+8)}{[(d-6) N+2 (d-12)]^2},
0,
\frac{12(N+3)(N+8)g^2 h^2}{p},
8g^3 h^5,
8dg  h^3\right)$
& 
2 for $N > N_\mathrm{c}$,\newline
1 for $N < N_\mathrm{c}$
&
O($N_\phi$) fixed point, critical for $2 \leq d<4$ and $N \leq N_\mathrm{c}$\\
IX&
$\left( \frac{(4-d)(N-1)}{4-d+N(d-7)},\infty, 3w, (N-4)w, \frac{81N^3(N+3)j^2}{m}, -\frac{27N^5j^3}{[8-2d+(d-6)N]^2}, -\frac{27N^3j}{k^2}\right)$
&
1 for $N > N_\mathrm{c}$,\newline
2 for $N < N_\mathrm{c}$
&
Cubic fixed point, critical for $2<d<4$ and $N > N_\mathrm{c}$\\
X&$\left( \frac{4-d}{d-10},\infty,0, \frac{24(4-d)}{(d-10)^2}, \frac{972(N+3)}{d-8}t^2, \frac{5832}{(d-6)^2}t^3, \frac{648}{(d-10)^2}t \right)$&
2
&
Decoupled Ising fixed point, critical for $d=2$ and $N \geq 3$\\[3pt]
\hline
\hline
\end{tabular*}
\label{tab:afm-pm-FP}
\end{table*}

In the mean-field phase diagram in Fig.~\ref{fig:mean-field-PD}(a,b), the transition between the symmetric and antiferromagnetic phases becomes continuous for sufficiently large tensor-field mass $r_T$. In this regime, the antiferromagnetic order parameter continuously develops a finite expectation value, which in turn induces a finite expectation value of the tensor field as a secondary order parameter.
This motivates studying the limit $r_T\rightarrow\infty$ within the perturbative RG framework. In this limit, the theory in Eq.~\eqref{eq:full-Hamiltonian}, together with the associated flow equations, simplifies considerably. More generally, for sufficiently large $r_T$, the tensor field can be integrated out, leaving an effective theory for the vector field with cubic anisotropy, as given in Eq.~\eqref{eq:vector-part}. We therefore expect the critical behavior in this regime to belong to the universality class of the cubic vector model.

To explicitly confirm this expectation within our RG approach, in which fluctuations of the vector and tensor fields are treated on equal footing, we consider the full model in Eq.~\eqref{eq:full-Hamiltonian} and take the limit $r_T\to\infty$ directly at the level of the flow equations, Eqs.~\eqref{eq:eta-functions-1}--\eqref{eq:lambda-beta-func}. Specifically, we neglect all terms proportional to $1/(1+r_T)$ or higher powers thereof, reducing the fixed-point equations to a set of six equations for the remaining parameters $r_{\phi\star}$, $u_{\phi\star}$, $v_{\phi\star}$, $u_{T\star}$, $\omega_\star^2$, and $\lambda_\star^2$.%
\footnote{Equivalently, one may introduce the compactified variable $y_T\coloneqq \tanh r_T$, for which $\beta_{y_T}=(1-y_T^2)\beta_{r_T}\bigr|_{r_T=r_T(y_T)}$, and then consider the fixed point at $y_{T\star}=1$.}
Importantly, in this limit, the flow equations for the vector-sector parameters $r_\phi$, $u_\phi$, and $v_\phi$ become independent of the tensor-sector parameters $u_T$ and $\omega$, as well as of the vector-tensor coupling $\lambda$. We have verified that the resulting fixed-point equations for $r_{\phi\star}$, $u_{\phi\star}$, and $v_{\phi\star}$ precisely coincide with those of the vector model with cubic anisotropy~\cite{pelissetto02}. The fixed-point values of $u_{T\star}$, $\omega_\star^2$, and $\lambda_\star^2$ can then be determined in a second step by substituting the fixed-point values of the vector-sector couplings into the remaining flow equations.

\paragraph{Above six dimensions.}

For $d>d_\mathrm{c}=6$, the system lies above the upper critical dimension, and the critical fixed point is the Gaussian fixed point, $(r_{\phi\star},r_{T\star},u_{\phi\star},v_{\phi\star},u_{T\star},\omega_\star^2,\lambda_\star^2)=(0,\infty,0,0,0,0,0)$, with a single relevant direction associated with $r_\phi$.

\paragraph{Between four and six dimensions.}

For $4<d<6$, the critical fixed point develops finite values in the tensor sector, $u_{T\star}=3(N_\phi+3)(6-d)^2 f^2/(d-8)$, $\omega^2_\star=(6-d)f^3$, and a finite vector-tensor coupling, $\lambda^2_\star=(6-d)f$. Here, $f \equiv f(d,a)$ is a function that depends only on dimension $d$ and the momentum-distribution parameter $a$. Its explicit form is given in Appendix~\ref{app:placeholders}.
In the vector sector, the critical fixed point remains noninteracting, $r_{\phi\star} = u_{\phi\star} = v_{\phi\star} = 0$, for all $d \geq 4$.
This can be understood from the fact that $d=4$ represents the upper critical dimension of the vector-only Hamiltonian in Eq.~\eqref{eq:vector-part}.
We therefore refer to this fixed point as ``Gaussian$'$ fixed point.''

\paragraph{Between two and four dimensions.}

For $2<d<4$, the critical fixed point becomes interacting in both the vector and tensor sectors. 
The character of the critical fixed point crucially depends on the value of $N_\phi$, see Table~\ref{tab:afm-pm-FP}.
Specifically, for $N_\phi < N_\mathrm{c}$, the critical fixed point is characterized by $v_{\phi\star} = 0$, indicating emergent O($N_\phi$) symmetry in the vector sector at criticality.
For $N_\phi > N_\mathrm{c}$, the cubic anisotropy in the vector sector becomes RG relevant, and the critical fixed point is characterized by $v_{\phi\star} \neq 0$.
This behavior is consistent with the expectation: For $r_T \to \infty$, we expect that the critical fluctuations in the vector sector can be understood in terms of the cubic vector-only model defined in Eq.~\eqref{eq:vector-part}, without the vector-tensor coupling. This model has previously been studied extensively within an $\epsilon$ expansion about  $d=4$~\cite{aharony73, folk00, carmona00, pelissetto02, calabrese02b}, finding $N_\mathrm{c} = 4 - 2(4-d) + \mathcal O((4-d)^2)$~\cite{kleinert95}. The leading term, corresponding to the one-loop order, precisely agrees with our result. Higher-loop corrections, however, tend to reduce $N_\mathrm{c}$. In fact, estimates from high-order perturbative expansions~\cite{carmona00, pelissetto02, calabrese02b}, recent conformal bootstrap calculations~\cite{rong18,chester21,rong23}, as well as state-of-the art Monte Carlo simulations~\cite{hasenbusch11,hasenbusch23,hasenbusch24}, place the value for $N_\mathrm{c}$ slightly below 3 when $d=3$, rendering the cubic anisotropy RG relevant at the antiferromagnetic-to-paramagnetic transition in the 3D case.

To further corroborate the scenario, we explicitly confirm the equivalence of the critical behavior of our coupled vector-tensor model in the limit $r_T \to \infty$ with the expectation from the vector-only model.
For $N_\phi < N_\mathrm{c}$, where the cubic anisotropy is irrelevant, we obtain the critical exponent $\nu$, which characterizes the divergence of the correlation length, $\xi \propto (r_\phi-r_{\phi,\mathrm c})^{-\nu}$, at the critical fixed point (denoted as ``VIII'' in Table~\ref{tab:afm-pm-FP}) as
\begin{align}
\label{eq:nu-O(N)}
\nu^\text{WF}=\frac{1}{2}+\frac{N_\phi+2}{4(N_\phi+8)}(4-d) + \mathcal O((4-d)^2)\,.
\end{align}
For the order-parameter anomalous dimension $\eta_\phi$, which characterizes the critical two-point correlator, $\langle\phi(x)\phi(0)\rangle \propto 1/|x|^{d-2+\eta_\phi}$, we find
\begin{align}
\eta_\phi^\text{WF} = \mathcal O((4-d)^2)\,.
\end{align}
These results precisely agree with the established one-loop results for the Wilson-Fisher (WF) O($N_\phi$) model~\cite{herbut07}.
It is further instructive to consider the tensor anomalous dimension $\eta_{T}$. At the O($N_\phi$) fixed point, we find
\begin{equation}
\label{eq:eta-T-O(N)}
\eta_T^\text{WF}=6-d - \frac{4}{N_\phi+8}(4-d) + \mathcal O((d-4)^2)\,.
\end{equation}
In our model, Eq.~\eqref{eq:full-Hamiltonian}, the tensor degree of freedom is treated as an independent coarse-grained field coupled to a bilinear of the vector order parameter. We denote the scaling dimension of this tensor field by $\Delta_\psi$, to distinguish it from the scaling dimension $\Delta_t$ of the tensor bilinear in the conventional O($N_\phi$) vector model (denoted by ``$\Delta_T$’’ in Ref.~\cite{kos14}). The latter is related to the scaling dimension of the tensor field in our model according to $\Delta_t=d-\Delta_\psi$. The scaling dimension $\Delta_\psi$, in turn, is related to the tensor anomalous dimension $\eta_T$ through $\Delta_\psi=(d-2+\eta_T)/2$.
Consequently, the tensor anomalous dimension $\eta_T$, together with the correlation-length exponent $\nu$, determines the crossover exponent $\phi_T=(d-\Delta_t)\nu = \Delta_\psi \nu$~\cite{pelissetto02, calabrese02a,gracey02,kos14}. This crossover exponent governs the shift of the critical point upon introducing a small cubic anisotropy, $r_{\phi,\mathrm{c}}(\delta)-r_{\phi,\mathrm{c}}(0)\propto\delta^{1/\phi_T}$, where $\delta$ parametrizes the anisotropy~\cite{riedel69,hikami74}. From Eqs.~\eqref{eq:nu-O(N)} and \eqref{eq:eta-T-O(N)}, we obtain
\begin{align}
\phi_T^\text{WF}
= 1 + \frac{N_\phi}{2(N_\phi + 8)} (4-d) + \mathcal O((4-d)^2)\,,
\end{align}
which again precisely agrees with the leading-order result for the conventional O($N_\phi$) universality class near and below four dimensions~\cite{fisher72,wegner72,calabrese02a,gracey02}.
For $N_\phi>N_\mathrm{c}$, where the cubic anisotropy is relevant, we obtain the correlation-length exponent at the critical fixed point (denoted as ``IX'' in Table~\ref{tab:afm-pm-FP})
\begin{align}
\nu^\text{cubic} = \frac{1}{2}+\frac{N_\phi - 1}{6 N_\phi}(4-d) + \mathcal O((4-d)^2)\,,
\end{align}
and the order-parameter anomalous dimension
\begin{align}
\eta_\phi^\text{cubic} = \mathcal O((4-d)^2)\,.
\end{align}
Both exponents precisely agree with those of the cubic fixed point near and below four dimensions~\cite{aharony73, pelissetto02}.
For the tensor anomalous dimension, we find at the cubic fixed point
\begin{align}
\eta_T^\text{cubic} =  6 - d - \frac{4}{3N_\phi}(4-d) + \mathcal O((4-d)^2)\,,
\end{align}
which gives the crossover exponent governing the shift of the critical point upon introducing an off-diagonal anisotropy $\propto \phi_a \phi_b$ with $a \neq b$,
\begin{align}
\phi_T^\text{cubic} = 1+ \frac{N_\phi-2}{3N_\phi}(4-d) + \mathcal O((4-d)^2)\,.
\end{align}
The corresponding scaling dimension of the off-diagonal ($T_{2g}$) component of the symmetric tensor is $\Delta^\text{cubic}_t = 2 - (3N_\phi - 2)(4-d)/(3N_\phi) + \mathcal O((4-d)^2)$, in full agreement with the result obtained in the cubic vector-only model~\cite{bednyakov23}. Note that $\phi_T^\text{cubic}$ and $\phi_T^\text{WF}$ coincide when the two fixed points merge at $N_\phi = N_\mathrm{c} = 4 + \mathcal O(d-4)$, as expected. 

\paragraph{Two dimensions.}

Lowering the dimension below $d=3$ causes the cubic fixed point to move toward the decoupled Ising fixed point, located at $u_{\phi\star}=0$ and $v_{\phi\star}\neq0$, thereby further reducing the value of $N_\mathrm{c}$ above which the cubic anisotropy becomes relevant. In fact, an $\epsilon$ expansion about the lower critical dimension yields $N_\mathrm{c}=2+4(d-2)+\mathcal O((d-2)^2)$~\cite{pelcovits76,newman82}. As $d\to2$, the cubic fixed point is expected to merge with the decoupled Ising fixed point at $N_\phi=3$~\cite{ran25}.
Consequently, in $d=2$, we expect the direct antiferromagnetic-to-paramagnetic transition to be continuous, with critical behavior in the Berezinskii-Kosterlitz-Thouless universality class for $N_\phi=2$ and in the Ising universality class for $N_\phi\geq3$. In the latter case, the crossover exponent $\phi_T^\text{Ising}$ associated with an off-diagonal bilinear anisotropy $\propto\phi_a\phi_b$ with $a\neq b$ follows from the fact that the two fields in the bilinear originate from decoupled Ising copies. Thus, $\Delta_t^\text{off}=2\Delta_\phi$, where $\Delta_\phi=\eta^\text{Ising}/2$ is the scaling dimension of the vector field. It follows that, in $d=2$, the crossover exponent coincides with the susceptibility exponent, $\phi_T^\text{Ising}=(2-\eta^\text{Ising})\nu^\text{Ising}=\gamma^\text{Ising}$, where the last equality follows from hyperscaling. Finally, we note that for $N_\phi=3$, the presence of a marginal operator associated with the fixed-point collision gives rise to multiplicative logarithmic corrections to the leading power-law behavior.

\paragraph{Summary.}

Let us briefly summarize the behavior of the antiferromagnetic-to-paramagnetic transition as a function of $r_\phi$ at fixed ultraviolet value of $r_T$, focusing on $N_\phi=3$. For sufficiently large $r_T$, we expect the transition to be direct and continuous, without an intermediate spin-vestigial phase. For $d\geq4$, the critical behavior is governed by mean-field theory, with multiplicative logarithmic corrections in $d=4$. In $d=3$, the transition is governed by the cubic universality class, with critical exponents $1/\nu^\text{cubic}=1.406\,25(50)$, $\eta_\phi^\text{cubic}=0.037\,82(13)$~\cite{hasenbusch23}, and $\phi_T^\text{cubic}/\nu^\text{cubic}=1.801\,2(76)$~\cite{rong23}. Finally, in $d=2$, the transition is governed by the decoupled Ising fixed point, with $\nu^\text{Ising}=1$, $\eta_\phi^\text{Ising}=1/4$, and $\phi_T^\text{Ising}=7/4$, again with multiplicative logarithmic corrections.

\subsubsection{Paramagnetic-to-nematic transition (\texorpdfstring{$r_\phi\rightarrow\infty$}{(rPhi->infty)})}
\label{subsubsec:nem-to-PM}

For sufficiently large $r_\phi$, the mean-field phase diagram in Figs.~\ref{fig:mean-field-PD}(a) and \ref{fig:mean-field-PD}(b) shows a direct first-order transition between the paramagnetic and spin nematic phases as a function of $r_T$.
To investigate the effects of fluctuations on the nature of this transition, we analyze the flow equations of the full model, Eqs.~\eqref{eq:eta-functions-1}--\eqref{eq:lambda-beta-func}, in the limit $r_\phi\rightarrow\infty$.
In analogy to the paramagnetic-to-antiferromagnetic transition studied above, we neglect all terms proportional to $1/(1+r_\phi)$ or higher powers thereof, reducing the fixed-point equations to a set of six equations for the remaining parameters $r_{T\star}$, $u_{\phi\star}$, $v_{\phi\star}$, $u_{T\star}$, $\omega_\star^2$, and $\lambda_\star^2$. Importantly, in this limit, the fixed-point value of the vector-tensor coupling vanishes, $\lambda_\star = 0$, such that the fixed-point equations for the remaining parameters $r_{T\star}$, $u_{T\star}$, and $\omega_\star^2$ in the tensor sector reduce to those of the four-state Potts model~\cite{zia75, amit76}.

\paragraph{Above six dimensions.}

For $d>6$, the Gaussian fixed point at $(r_{T\star},u_{T\star},\omega_\star^2)=(0,0,0)$ is critical with a single relevant direction associated with $r_T$. 
We interpret it to describe critical fluctuations around the metastable state in the vicinity of the first-order transition~\cite{wu82}, cf.~Appendix~\ref{app:metastable}.

\paragraph{Between two and six dimensions.}

For $d=6-\epsilon$ and $0<\epsilon\ll1$, the critical fixed point is located at $(r_{T\star},u_{T\star},\omega^2_\star) = (-\epsilon/2, -3\epsilon^2(N_T+3)/2, -\epsilon)$, representing another instance of an ``imaginary CFT''~\cite{wiese24}.
The fact that the fixed-point value for the cubic coupling $\omega_\star$ is complex suggests the absence of any critical fluctuations about the metastable state.
We note that the $r_\phi=0$ counterpart of this fixed point was identified in the discussion of possible multicritical behavior in Sec.~\ref{subsubsec:triple-point} as Fixed Point~V in Table~\ref{tab:epsilon-FP}, where it is interpreted as a decoupled Potts fixed point, with additional relevant directions associated with $r_\phi$ and $\lambda$.
To gain insight into the physics away from the upper critical dimension, we track the location of the critical fixed point numerically to lower dimensions. We find that the fixed point remains complex at $d=5$, but reemerges into the real-coupling space at a critical dimension $d_\mathrm{c,Potts} < 5$. Within our one-loop approximation, this occurs at $d_\mathrm{c,Potts}\simeq 4.09$; however, higher-loop corrections are expected to shift the critical dimension to $d_\mathrm{c,Potts}=2$~\cite{wu82,wiese24}, implying that the transition remains discontinuous for all $d>2$.

\paragraph{Two dimensions.}

In two dimensions, the critical fixed point has reemerged into the real-coupling space, and the transition becomes continuous. The critical exponents are $\nu^\text{Potts}=2/3$ and $\eta_T^\text{Potts}=1/4$~\cite{baxter73,wu82,gorbenko18b}, with multiplicative logarithmic corrections to the leading power-law behavior~\cite{nauenberg80,cardy80}.

\subsubsection{Nematic-to-antiferromagnetic transition (\texorpdfstring{$r_T\rightarrow-\infty$}{rT->-infty})}
\label{subsubsec:nem-to-AFM}

Below the triple point A, the mean-field phase diagram in Figs.~\ref{fig:mean-field-PD}(a) and \ref{fig:mean-field-PD}(b) exhibits a continuous transition between the spin-nematic and antiferromagnetic phases as a function of $r_\phi$ for fixed $r_T<r_T\bigr|_\mathrm{A}$. This transition occurs in the presence of strong nematic order, corresponding to the limit $r_T\to-\infty$. However, our flow equations, Eqs.~\eqref{eq:eta-functions-1}--\eqref{eq:lambda-beta-func}, were derived for vanishing or small order parameters and therefore cannot be directly applied in this regime.
To analyze the critical fluctuations of the antiferromagnetic order parameter, we instead treat the strong nematic order as a fixed background and neglect fluctuations of the nematic order parameter. This approximation is expected to become asymptotically valid as $r_T\to-\infty$.
For $\omega>0$ in Eq.~\eqref{eq:nematic-part}, the nematic order parameter $\boldsymbol{\psi}$ is then restricted to one of the four cubic body diagonals shown in Fig.~\ref{fig:minima-sketches}(b), which can be parametrized as
\begin{equation}
\label{eq:rt-minus-infty-nem-OP}
\boldsymbol{\psi}=\frac{\psi_0}{\sqrt{3}}(\eta_1,\eta_2,\eta_3)
\quad \text{where $\eta_a = \pm 1$ and $\eta_1\eta_2\eta_3 = -1$}\,.
\end{equation}
The nematic background affects the vector sector through the vector-tensor interaction, Eq.~\eqref{eq:simplified-interacting-term}, which acts as an effective off-diagonal contribution to the vector-field mass matrix,
\begin{equation}
    \label{eq:rt-minus-infty-interaction}
    \mathcal{H}_{\phi\psi}^{\mathrm{eff}}=\frac{\lambda\psi_0}{2\sqrt{3}}  \int \rmd^dx \, \sum_{a,b,c}|\epsilon^{abc}|\eta_a \phi_b\phi_c\,.
\end{equation}
Depending on the sign structure of $(\eta_a)$, this contribution favors vector configurations along one of the four body diagonals. For example, for $\eta_1=\eta_2=\eta_3=-1$, configurations with $\phi_x=\phi_y=\phi_z$ are favored, corresponding to the $\pm[111]$ direction. Thus, the eight degenerate minima of the vector sector shown in Fig.~\ref{fig:minima-sketches}(a) are reduced to the two minima shown in Fig.~\ref{fig:minima-sketches}(c).
In the long-wavelength limit, the resulting effective theory describes spontaneous $\mathbb{Z}_2$ symmetry breaking, suggesting that the nematic-to-antiferromagnetic transition is continuous and belongs to the Ising universality class. The corresponding critical exponents are $\nu^\text{Ising}=1$ and $\eta^\text{Ising}_T=1/4$ in $d=2$, while in $d=3$ they are $\nu^\text{Ising}=0.629\,970\,97(12)$ and $\eta^\text{Ising}_T=0.036\,297\,612(48)$~\cite{chang25}.
In Appendix~\ref{app:rt-minus-infty}, we demonstrate that the vector model with the effective off-diagonal contribution to the mass matrix in Eq.~\eqref{eq:rt-minus-infty-interaction} can be explicitly mapped onto a $\mathbb Z_2$ Ising field theory for the modes along the selected diagonal, thereby confirming the Ising critical behavior of the nematic-to-antiferromagnetic transition.

\subsection{Integration of RG flow}
\label{subsec:summary}

The fixed-point analysis of the previous subsection suggests a flow diagram sketched schematically for $2<d<4$ in Fig.~\ref{fig:mean-field-PD}(c).
The multicritical and four-state Potts fixed points are complex, implying first-order transitions at the triple point and between the paramagnetic and spin-nematic phases, respectively. In the limit $r_\phi\rightarrow\infty$, a critical O($N_\phi$) or cubic fixed point, depending on $N_\phi$ and $d$, describes a continuous paramagnetic-to-antiferromagnetic transition. Similarly, the continuous nematic-to-antiferromagnetic transition is governed by an Ising fixed point in the limit $r_\phi\rightarrow-\infty$. Since our approach retains the vector and tensor degrees of freedom as independent coarse-grained fields, we can test this picture by explicitly integrating the RG flow equations for different ultraviolet starting values.

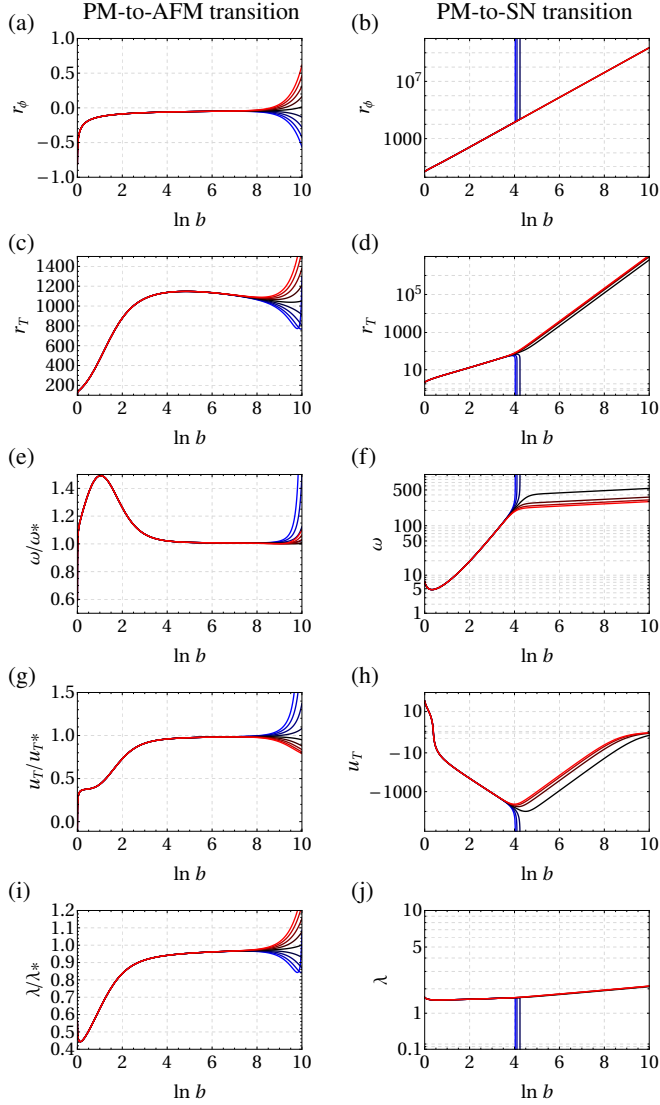
\begin{figure}[tb!]
\centering
\begin{overpic}[width=\columnwidth]{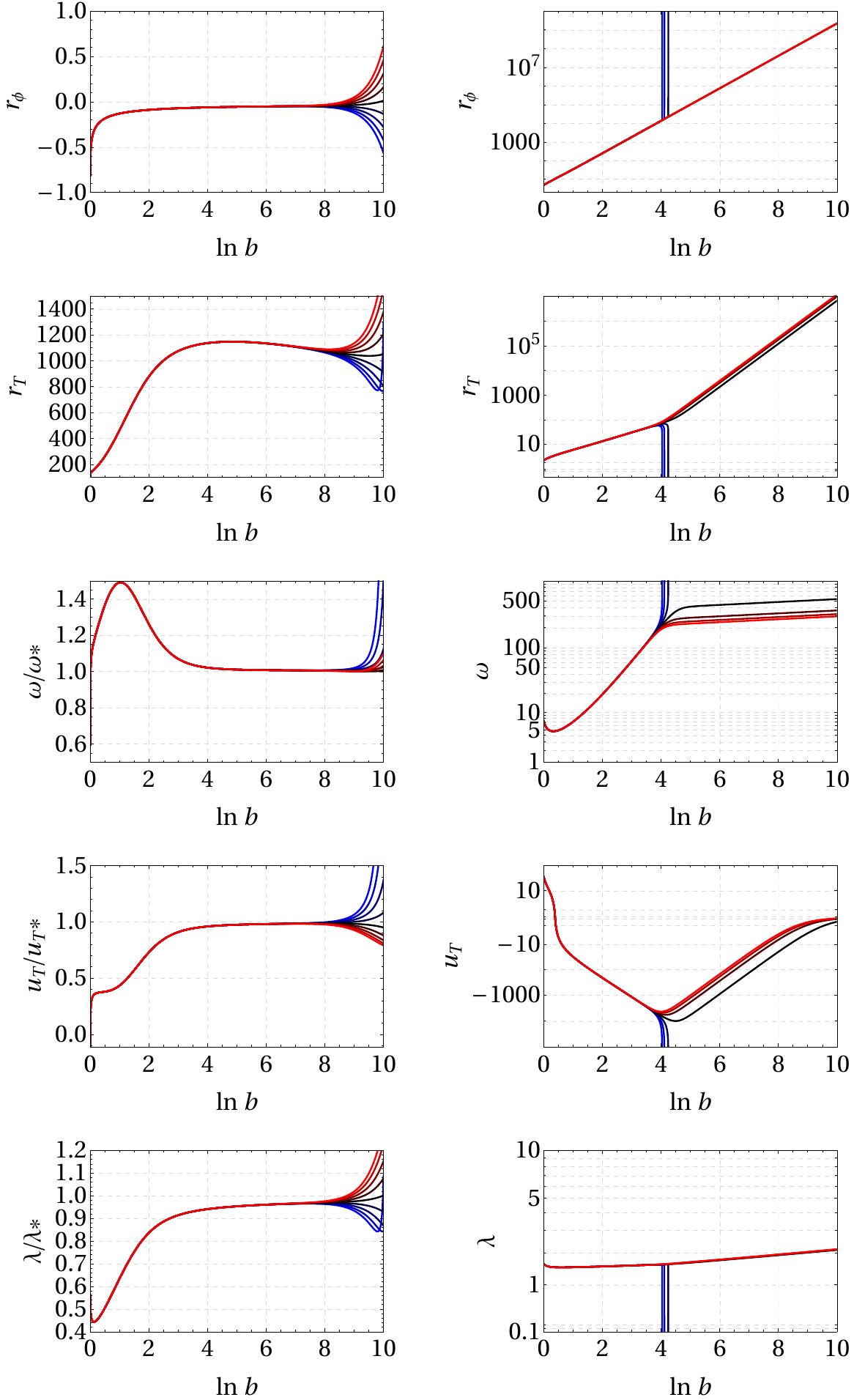}
\put(0,100){(a)}
\put(7,101){PM-to-AFM transition}
\put(0,79.5){(c)}
\put(0,59.5){(e)}
\put(0,39){(g)}
\put(0,19){(i)}
\put(40,101){PM-to-SN transition}
\put(32,100){(b)}
\put(32,79.5){(d)}
\put(32,59.5){(f)}
\put(32,39){(h)}
\put(32,19){(j)}
\end{overpic}
\caption{%
Left column: RG flow of 
(a)~the vector mass $r_\phi$,
(c)~the tensor mass $r_T$,
(e)~the cubic tensor self-coupling $\omega$,
(g)~the quartic tensor self-coupling $u_T$, and
(i)~the vector-tensor coupling $\lambda$,
near the paramagnetic-to-antiferromagnetic (PM-to-AFM) transition at $d=3.9$, shown as a function of the RG time $\ln b$ for different ultraviolet starting values of the vector mass above (red curves), close to (black curves), and below (blue curves) $r_{\phi,\mathrm c}$.
We use $(r_{\phi},r_T)\bigr|_{\ln b = 0} = (r_{\phi,\mathrm{c}} + \delta r_\phi,100)$ and $(u_\phi,v_\phi,u_T,\omega,\lambda)\bigr|_{\ln b =0}=(24,12,36,7.2,1.6)S_d/(2\pi)^d$, with $r_{\phi,\mathrm c} \approx -0.8096679527$ and $\delta r_\phi \in \{\sim 0^+,\pm 3, \pm 6, \pm 9, \pm 12\} \times 10^{-10}$.
%
The couplings in (e,g,i) are normalized to their values at the O(3) fixed point (cf.\ Table~\ref{tab:afm-pm-FP}).
Right column: Same as left column, but near the paramagnetic-to-nematic (PM-to-SN) transition at $d=5.9$, for different ultraviolet starting values of the tensor mass above (red curves), close to (black curves), and below (blue curves) $r_{T,\mathrm{c}}$. Here, $(r_{\phi},r_T)\bigr|_{\ln b = 0} = (4, r_{T,\mathrm{c}} + \delta r_T)$, with $r_{T,\mathrm c} \approx 1.44112050908$ and $\delta r_T \in \{\sim 0^+, \pm 0.5, \pm 1.0, \pm 1.5\} \times 10^{-11}$.
%
%
}
\label{fig:flow-comparison}
\end{figure}

We fix $N_\phi = N_T = 3$ and first consider the paramagnetic-to-antiferromagnetic transition. The panels in the left column of Fig.~\ref{fig:flow-comparison} show the integrated RG flow for three different starting values of the vector mass $r_\phi$, chosen above, at, and below the critical value $r_{\phi,\mathrm c}$. We use $d=3.9$, where the O(3) and cubic fixed points are perturbatively accessible, cf.\ Table~\ref{tab:afm-pm-FP}. For $r_\phi>r_{\phi,\mathrm c}$ (red curves), the masses of both the vector and tensor fields increase toward the infrared, indicating that the flow approaches the symmetric regime corresponding to the paramagnetic phase. For $r_\phi<r_{\phi,\mathrm c}$ (blue curves), the vector mass $r_\phi$ becomes negative during the RG flow, indicating that the system enters the regime of spontaneous cubic and time-reversal symmetry breaking corresponding to the antiferromagnetic phase. We terminate the flow once $r_\phi$ approaches $-1$, since our RG equations were derived in the symmetric regime and cannot be applied reliably for large negative $r_\phi$.
For the fine-tuned value $r_\phi \simeq r_{\phi,\mathrm c}$ (black curves), the flow is attracted to the critical fixed point, which is the O(3) fixed point for small $4-d$. This is reflected in the plateaus of $\omega$, $\lambda$, and $u_T$ shown in Figs.~\ref{fig:flow-comparison}(e,g,i). Upon lowering the dimension, we expect qualitatively similar behavior, with the critical flow governed by the cubic fixed point in $d=3$~\cite{pelissetto02,calabrese02b,rong18,chester21,rong23,hasenbusch11,hasenbusch23,hasenbusch24} and by the decoupled Ising fixed point in $d=2$~\cite{ran25}. Thus, for all $d\geq2$, we expect the paramagnetic-to-antiferromagnetic transition to remain continuous for sufficiently large initial $r_T$.

We next turn to the paramagnetic-to-nematic transition. The panels in the right column of Fig.~\ref{fig:flow-comparison} show the integrated RG flow for three different starting values of the tensor mass $r_T$, chosen above, at, and below the critical value $r_{T,\mathrm c}$. Here, we use $d=5.9$, where the Potts fixed point is perturbatively accessible, cf.\ Sec.~\ref{subsubsec:nem-to-PM}. For $r_T>r_{T,\mathrm c}$ (red curves), the masses of both the vector and tensor fields increase toward the infrared, indicating that the flow approaches the symmetric regime corresponding to the paramagnetic phase. For $r_T<r_{T,\mathrm c}$ (blue curves), the tensor mass $r_T$ becomes negative during the RG flow, indicating that the system enters the regime of spontaneous cubic symmetry breaking without time-reversal symmetry breaking corresponding to the spin-nematic phase. As above, we terminate the flow once $r_T$ approaches $-1$, since our RG equations were derived in the symmetric regime and cannot be applied reliably for large negative $r_T$.
For the fine-tuned value $r_T \simeq r_{T,\mathrm c}$ (black curves), no plateau develops in the RG flow. Instead, the flow exhibits runaway behavior toward either the paramagnetic or the nematic regime, consistent with the absence of a critical Potts fixed point in the real-coupling space.
We therefore conclude that the paramagnetic-to-nematic transition is first order, without critical fluctuations about the metastable state. We expect the same behavior for all dimensions $2<d<6$. In $d=2$, by contrast, the Potts fixed point reemerges into the real-coupling plane~\cite{wu82,wiese24}, rendering the paramagnetic-to-nematic transition continuous.

\section{Conclusion}
\label{sec:conclusion}

In this work, we have developed a minimal continuum field theory describing the competition between primary dipolar and secondary quadrupolar orders.
We considered the cubic symmetry group $O_h\simeq S_4\times \mathbb{Z}_2$ and described the competing orders by independent coarse-grained fields, treating their fluctuations on equal footing. Under $O_h$, the tensor field decomposes into triplet and doublet irreducible representations that do not mix under the group action. We focused on the $T_{2g}$ triplet component, which is relevant to the competition between the $\mathbb Z_4$ spin-current density wave and triple-$\mathbf q$ antiferromagnetic order recently observed in numerical simulations of the extended Kitaev-Heisenberg model on the honeycomb lattice~\cite{francini24vestigial}, possibly realized in the candidate Kitaev material \NCTO~\cite{krueger23,francini24ferri,francini25ferri}.

Using mean-field and renormalization group (RG) analyses, we have investigated the resulting phases and phase transitions.
The phase diagram contains three phases: a fully symmetric phase corresponding to the high-temperature paramagnet; a magnetically ordered phase that breaks time-reversal and spin-rotational symmetries, corresponding to the triple-$\mathbf q$ antiferromagnet observed in \NCTO at low temperatures~\cite{krueger23}; and a spin-nematic vestigial phase, in which spin-rotational symmetry is spontaneously broken while time-reversal symmetry remains preserved. The latter phase may be realized in \NCTO\ at intermediate temperatures between 27\,K and 31\,K~\cite{chen21,francini24vestigial}.
In $d=2$, the transition from the high-temperature paramagnet to the spin-vestigial phase can be continuous, in which case the critical behavior belongs to the four-state Potts universality class.
In $d=3$, potentially relevant to extended Kitaev-Heisenberg models on hyperhoneycomb and stripyhoneycomb lattices~\cite{krueger20}, the paramagnetic-to-nematic transition is necessarily first order within our field-theory analysis.
The transition between the spin-nematic and antiferromagnetic phases is generically expected to be continuous in both $d=2$ and $d=3$, although an effective field theory alone cannot exclude a first-order transition. If continuous, the transition belongs to the Ising universality class.
Importantly, a direct transition from the high-temperature paramagnet to the low-temperature antiferromagnet, without an intervening spin-nematic phase, need not be first order, contrary to generic expectations~\cite{golubovic88, fernandes12, fernandes19, hecker23}. In our model, this transition becomes continuous when the tensor mass $r_T$ at the transition point is sufficiently large. By explicitly integrating the RG flow, we have verified that this conclusion persists beyond mean-field theory when order-parameter fluctuations are included.
In the vicinity of the triple point, however, where the paramagnetic, spin-nematic, and antiferromagnetic phases meet, the direct paramagnetic-to-antiferromagnetic transition is necessarily first order within our field-theory analysis. The first-order nature persists along a path in parameter space passing through the triple point. This conclusion is based on the observation that the multicritical fixed point is located at complex coupling for $d<6$, at least within the one-loop approximation considered here. It would be interesting to test this prediction using more accurate methods, for example by extending the present analysis to higher-loop order or by employing a nonperturbative approach analogous to those developed in Refs.~\cite{zinati18, sanchezvillalobos23, wiese24}.

We expect analogous behavior for competing primary dipolar and secondary quadrupolar orders when the $E_g$ doublet component of the tensor field is energetically favored.
In $d=2$, this situation can be realized in the extended Kitaev-Heisenberg model on the honeycomb lattice in the regime where zigzag antiferromagnetic order is stabilized at low temperatures~\cite{francini24vestigial}.
In this case, a continuous paramagnetic-to-nematic transition belongs to the three-state Potts universality class, while the nematic-to-antiferromagnetic transition, if continuous, is of Ising type. The direct paramagnetic-to-antiferromagnetic transition is first order near the triple point. These predictions are consistent with numerical simulations of the extended Kitaev-Heisenberg model on the honeycomb lattice~\cite{francini24vestigial}.
It would be interesting to determine whether a continuous paramagnetic-to-antiferromagnetic transition also becomes possible sufficiently far from the triple point when the $E_g$ doublet component of the tensor field is energetically favored.
In $d=3$, the paramagnet-to-nematic transition is necessarily discontinuous, as the three-state Potts fixed point becomes complex in the range $2.5 \lesssim d \lesssim 5.9$~\cite{wiese24,chester25}.
The nematic-to-antiferromagnetic transition, by contrast, can be continuous and would then again belong to the Ising universality class.
These predictions are consistent with the nature of the phase transitions into and out of the spin-nematic phase observed in numerical simulations of a nearest-neighbor spin model on the pyrochlore lattice~\cite{francini2025exactnematic}. 
%
%
Beyond the triple point, the simulations also reveal a direct transition from the paramagnetic phase to the primary $A_2 \oplus \psi_2$ phase, without an intervening spin-nematic phase. This transition appears to be continuous in the simulations, with critical behavior speculated to belong to the U(1) universality class. It would be interesting to test this scenario within a continuum-field-theory approach analogous to the one developed here.

Our approach differs from the large-$N$ methods commonly used to study vestigial orders~\cite{fernandes19,hecker24,palle26,obrien26vestigialnematicorderzero} in that we retain both the vector and tensor degrees of freedom as independent coarse-grained fields and treat their fluctuations on equal footing.
This formulation also allows us to investigate the case of a critical primary field and a noncritical composite field, relevant to the direct paramagnetic-to-antiferromagnetic transition observed in our model and in recent simulations~\cite{francini24vestigial,francini2025exactnematic}.
Recent work has emphasized the Fierz ambiguity inherent in conventional large-$N$ approaches~\cite{palle26}. A related ambiguity can arise in partially bosonized approaches: if the Hubbard-Stratonovich transformation is performed only once at the microscopic scale, different choices of the bosonization channel, related by Fierz transformations, can lead to different results within a given approximation~\cite{jaeckel03}. In this sense, our use of partial bosonization at the ultraviolet scale may also leave a residual dependence on the choice of bosonic representation.
For fermionic systems, this ambiguity can be overcome within a RG approach employing dynamical bosonization~\cite{jaeckel03}. In this approach, a Hubbard-Stratonovich transformation is performed after each RG step, allowing newly generated quartic interactions of the primary field to be absorbed into a redefinition of the composite field~\cite{gies02, pawlowski07, floerchinger09, janssen17a, moser25}.
It would therefore be interesting to study our continuum field theory using dynamical bosonization in future work. This would also provide a direct comparison with the results of unbiased large-$N$ approaches~\cite{palle26}.

Finally, we note that the effective field theory in Eq.~\eqref{eq:full-Hamiltonian} can be simulated directly using classical Monte Carlo methods after discretization on a $d$-dimensional lattice.
Such simulations would allow the phase diagram of the discretized theory to be mapped out numerically and would provide a direct test of our theoretical predictions for the nature of the various phase transitions in $d=2$ and $d=3$.

\begin{acknowledgments}
We thank Junchen Rong and Thomas Vojta for insightful discussions.
This work has been supported by the Deutsche Forschungsgemeinschaft through 
Project No.\ 247310070 (SFB 1143, A07), 
Project No.\ 390858490 (W\"urzburg-Dresden Cluster of Excellence \textit{ctd.qmat}, EXC 2147), and 
Project No.\ 411750675 (Emmy Noether program, JA2306/4-1).
\end{acknowledgments}

\section*{Data availability}

The data that support the findings of this article are openly
available~\cite{data-availability}.
%

\appendix

\section{Critical fluctuations about a metastable state}
\label{app:metastable}

\begin{figure}[tb!]
\centering
\begin{overpic}[width=\linewidth]{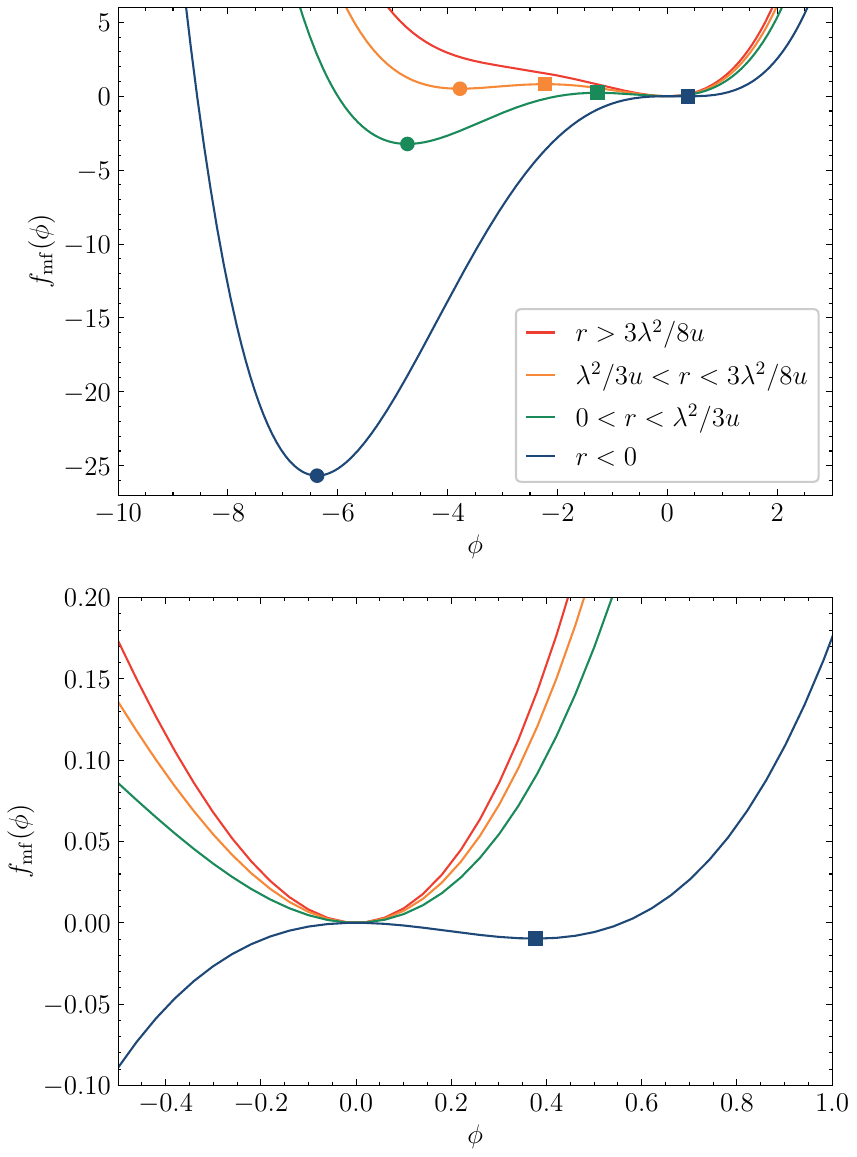} 
\put(0,100){(a)}
\put(0,49){(b)}
\end{overpic}
\caption{%
(a)~Schematic Landau mean-field potential of the $\phi^3$ field theory as a function of the order parameter $\phi$ for fixed $\lambda, u > 0$ and different values of $r$.
For $r>3\lambda^2/8u$ (red curve), the energy has a unique minimum at $\phi_0=0$.
For $\lambda^2/3u < r < 3\lambda^2/8u$ (orange curve), a local maximum at $\phi_+ < 0$ (orange square) and a local minimum at $\phi_- < \phi_+$ (orange dot) appear.
At $r = \lambda^2/3u$, the global minimum jumps from $\phi_0 = 0$ to $\phi_- < 0$ through a level crossing, signaling a first-order phase transition.
For $0 < r < \lambda^2/3u$ (green curve), the global minimum is at $\phi_- < 0$  (green dot), while $\phi_0 = 0$ remains a local minimum corresponding to a metastable state.
As $r \to 0^+$, the local maximum at $\phi_+<0$ approaches the local minimum at $\phi_0=0$. The resulting fluctuations about the metastable state are universal if the RG flow possesses a critical fixed point in the real-coupling space.
For $r<0$ (blue curve), $\phi_+ > 0$ (blue square) becomes the local minimum corresponding to the metastable state, while $\phi_0 = 0$ becomes a local maximum.
(b)~Zoom of the region around $\phi_0 = 0$. For $r<0$, $\phi_+$ (blue square) becomes a local minimum.
}
\label{fig:phi3-plot}
\end{figure}

In this appendix, we illustrate how critical fluctuations about a metastable state in the vicinity of a first-order transition can be described within an RG approach, as advertised in Secs.~\ref{subsubsec:triple-point} and \ref{subsubsec:nem-to-PM}.
As a toy model, we consider the simple $\phi^3$ field theory, which is related to ordinary percolation problems~\cite{priest-76-percolation,fisher78-phi3,gracey15-phi3,borinsky21-phi3}, with Hamiltonian
\begin{equation}
    \label{eq:phi3-model}
    \mathcal{H}=\int \rmd^dx \left[\frac{1}{2}(\nabla\phi)^2+ \frac{r}{2}\phi^2+\frac{\lambda}{3!}\phi^3 + \frac{u}{4!}\phi^4\right],
\end{equation}
where $\phi$ is a single-component real scalar field. We take the couplings to be positive, $\lambda,u>0$, and tune the system through the different regimes by varying the mass parameter $r$.
The Landau mean-field potential $f_\text{mf}(\phi)$ of this model has up to three stationary points, shown in Fig.~\ref{fig:phi3-plot}. Depending on $r$, these correspond to unstable, metastable, or stable states. The stationary points are located at
\begin{align}
\label{eq:phi3-stationary-points}
    \phi_0& =0\,, & 
    \phi_{\pm} & =\frac{3\lambda}{2u}\left[-1\pm\sqrt{1-\frac{8ru}{3\lambda^2}}\right]\,.
\end{align}
For $r>3\lambda^2/(8u)$, the solutions $\phi_\pm$ are complex, leaving a single global minimum at $\phi_0=0$ and hence a unique stable state.
For $r\leq3\lambda^2/(8u)$, the solutions $\phi_\pm$ become real.
In the range $\lambda^2/(3u)<r\leq3\lambda^2/(8u)$, the global minimum remains at $\phi_0=0$, while a local minimum develops at $\phi_-<0$. The solution $\phi_+>\phi_-$ is a local maximum and therefore an unstable state.
For $0<r<\lambda^2/(3u)$, the global minimum shifts from $\phi_0=0$ to $\phi_-<0$, while $\phi_0=0$ becomes metastable. This marks a first-order transition with a finite jump of the order parameter.
As $r\to0^+$, the local maximum at $\phi_+<0$ approaches the local minimum at $\phi_0=0$. At $r=0$, the curvature at $\phi_+ = \phi_0=0$ vanishes, so that the correlation length about the metastable state diverges. The resulting critical fluctuations can therefore be studied using RG methods. In what follows, we argue that these fluctuations exhibit universal behavior whenever the RG flow possesses a critical fixed point in the real-coupling space.
For $r<0$, the solution $\phi_+>0$ becomes metastable, while $\phi_0=0$ is now an unstable state.

Mean-field theory is expected to accurately describe the critical behavior of fluctuations about the metastable state in dimensions $d$ above the upper critical dimension $d_\mathrm{c}$.
The engineering scaling dimension of the cubic coupling in the $\phi^3$ theory is $[\lambda]=(6-d)/2$.
As in the full model described in the main text [Eq.~\eqref{eq:full-Hamiltonian}], the upper critical dimension is therefore $d_\mathrm{c}=6$.
In the vicinity of the metastable state, the leading nonlinearity is $\lambda\phi^3$ rather than $u\phi^4$. Consequently, the exponents $\alpha$, $\beta$, and $\delta$, which are sensitive to the leading nonlinearity, differ from those of conventional $\phi^4$ criticality already at the mean-field level.
For example, expanding the position of the stationary point $\phi_+$ in Eq.~\eqref{eq:phi3-stationary-points} to leading order in $r$ gives $\phi_+=2(-r)/\lambda$ for $r<0$, yielding $\beta=1$. Upon applying a small conjugate field $h$ that couples linearly to $\phi$, the stationary point at $\phi_+=0$ for $r=0$ shifts to $\phi_+=\sqrt{2h/\lambda}$, yielding $\delta=2$. The susceptibility $\chi=\partial\phi_+/(\partial h)\bigr|_{h\to0}$ of the metastable state scales as $\chi=1/|r|$, implying $\gamma=1$. Finally, assuming $r\propto T-T_\mathrm{meta,c}$, where $T_\mathrm{meta,c}$ denotes the temperature at which the correlation length in the metastable state diverges, the heat capacity scales as
$C\propto-\partial^2 f_\text{meta}/\partial r^2=4|r|/\lambda^2$ for $r<0$, corresponding to $\alpha=-1$. Here, $f_\text{meta}$ denotes the mean-field free energy of the metastable state.
At the mean-field level, the correlation function in the metastable state has the same scaling form as in conventional critical $\phi^4$ theory, with anomalous dimension $\eta=0$ and correlation-length exponent $\nu=1/2$.

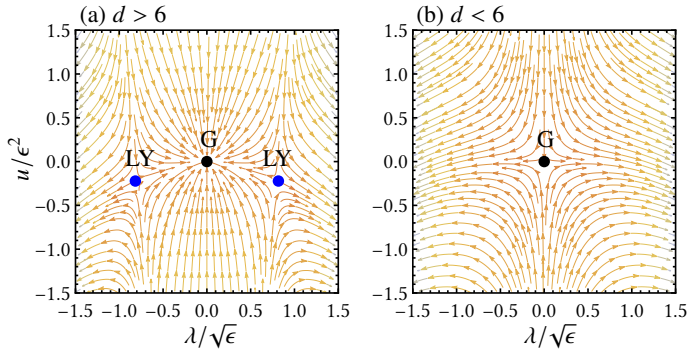
\begin{figure}[tb!]
\centering
\begin{overpic}[width=\linewidth]{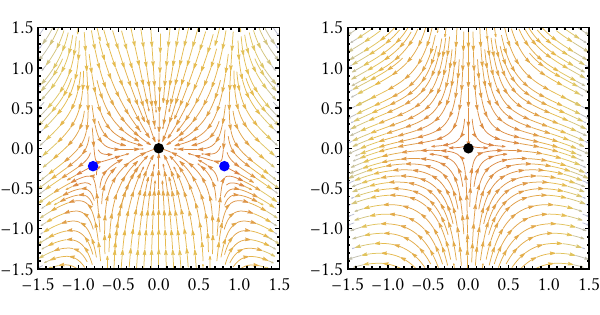}
\put(-4,27){\rotatebox[]{90}{$u/\epsilon^2$}}
\put(23,0){$\lambda/\sqrt{\epsilon}$}
\put(75,0){$\lambda/\sqrt{\epsilon}$}
\put(25.5,30){G}
\put(77,30){G}
\put(14,27){LY}
\put(35,27){LY}
\put(7,49){(a) $d>6$}
\put(58.5,49){(b) $d<6$}
\end{overpic}
\caption{%
Renormalization group flow in the $\phi^3$ theory on the critical surface $r=r_\star$ as a function of the cubic coupling $\lambda$ and quartic coupling $u$, near the upper critical dimension $d_\mathrm{c} = 6$.
(a)~For $d>6$, the Gaussian fixed point G (black dot) is critical, while the Lee-Yang fixed points LY (blue dots) are unstable. Critical fluctuations about the metastable state are universal and governed by the Gaussian fixed point.
(b)~For $d<6$, the Lee-Yang fixed points move to purely imaginary $\lambda_\star$, while the Gaussian fixed point (black dot) becomes unstable. The resulting runaway flow indicates the absence of universal critical fluctuations about the metastable state.
}
\label{fig:phi3-flow}
\end{figure}

For $d<d_\mathrm{c}=6$, fluctuations about the metastable state can be described within an RG framework.
Employing the same momentum-shell approach in fixed dimension as in the main text, we obtain the eta function
\begin{align}
\eta=\frac{-2+d\left(1+r\right) + \left(1-2a\right)^2\left[-6+d(1+r)\right]}{4d(1+r)^4}\lambda^2\,,
\end{align}
and the beta functions
\begin{align}
\beta_r&=\left(2-\eta\right)r+\frac{u}{2(1+r)}-\frac{\lambda^2}{2(1+r)^2}\,, \label{eq:beta-r-phi3} \displaybreak[0]\\
\beta_\lambda&=\left(\frac{6-d-3\eta}{2} -\frac{3u}{2(1+r)^2}+\frac{\lambda^2}{(1+r)^3}\right)\lambda\,, \label{eq:beta-lambda-phi3}\displaybreak[0]\\
\beta_u&=\left(4-d-2\eta\right)u -\frac{3u^2}{2(1+r)^2} +\frac{6\lambda^2u}{(1+r)^3} -\frac{\lambda^4}{(1+r)^4}\,, \label{eq:beta-u-phi3}
\end{align}
where $a$ parametrizes the momentum splitting in one of the diagrams renormalizing $r$, with 
$0\leq a \leq 1$.
In the above equations, we have rescaled the parameters according to
\begin{equation}
    \label{eq:rescaling-variable-phi3}
    \frac{r}{\Lambda^2}\mapsto r, \quad  \frac{\lambda^2S_d\Lambda^{d-6}}{(2\pi)^d} \mapsto \lambda^2    , \quad \frac{uS_d\Lambda^{4-d}}{(2\pi)^d} \mapsto u\,.
\end{equation}
This rescaling is analogous to that used in the main text [Eqs.~\eqref{eq:rescaling-1}--\eqref{eq:rescaling-3}].
For small $\epsilon=6-d$, the flow equations~\eqref{eq:beta-r-phi3}--\eqref{eq:beta-u-phi3} admit three fixed points,
\begin{align}
\label{eq:phi3-fixed-points}
\text{G}&:& (r_\star,\lambda^2_\star,u_\star)& =(0,0,0)\,, \displaybreak[0]\\
\text{WF}&:& (r_\star,\lambda^2_\star,u_\star) & = \left(\frac{1}{2}-\frac{3}{8}\epsilon,0,-3+3\epsilon\right) + \mathcal{O}(\epsilon^2)\,, \displaybreak[0]\\ 
\text{LY}&:& (r_\star,\lambda^2_\star,u_\star) & = \left(-\frac{\epsilon}{6},-\frac{2}{3}\epsilon,-\frac{2}{9}\epsilon^2\right) + \mathcal{O}(\epsilon^2)\,.
\end{align}
G is the Gaussian fixed point, while WF corresponds to the Wilson-Fisher fixed point of the $\phi^4$ theory for $\lambda=0$. The nontrivial fixed points LY at nonzero $\lambda_\star=\pm\sqrt{\lambda_\star^2}$ are specific to the $\phi^3$ field theory.
For $d>6$, all fixed points are real, and the Gaussian fixed point G is critical, with a single RG-relevant direction associated with $r$, see Fig.~\ref{fig:phi3-flow}(a). The corresponding critical exponents are $\nu=1/2$ and $\eta=0$. In the limit $d\to6^+$, the hyperscaling relations yield $\alpha=-1$, $\beta=1$, $\gamma=1$, and $\delta=2$, in precise agreement with the mean-field exponents discussed above. We therefore associate the Gaussian fixed point with universal critical fluctuations about the metastable state in the $\phi^3$ field theory for $d\geq 6$.
For $d<6$, the Gaussian fixed point develops an additional RG-relevant direction, and the nontrivial fixed points LY become critical.
However, at the LY fixed points, the coupling $\lambda^2_\star<0$, implying that the cubic coupling $\lambda_\star$ itself is purely imaginary. This leads to a runaway flow on the critical surface, see Fig.~\ref{fig:phi3-flow}(b).
The critical exponents associated with the imaginary fixed points LY are nevertheless real,
\begin{align}
\nu & = \frac{1}{2}+\frac{5}{36}\epsilon + \mathcal O(\epsilon^2)\,, &
\eta & = -\frac{\epsilon}{9}  + \mathcal O(\epsilon^2)\,.
\end{align}
Assuming hyperscaling, the remaining critical exponents are $\alpha=-1-\epsilon/3+\mathcal{O}(\epsilon^2)$, $\beta=1+\mathcal{O}(\epsilon^2)$, $\gamma=1+\epsilon/3+\mathcal{O}(\epsilon^2)$, and $\delta=2+\epsilon/3+\mathcal{O}(\epsilon^2)$.
Because the LY fixed points are located at purely imaginary coupling $\lambda_\star$, these exponents are not accessible in a unitary theory. The fixed points instead describe a nonunitary conformal field theory, namely the Lee-Yang (LY) theory~\cite{fisher78-phi3, gorbenko18a, wiese24}.
For the unitary model in $d<6$, the absence of a critical fixed point in the real-coupling space suggests that fluctuations about the metastable state are nonuniversal.

\section{Auxiliary functions used in the fixed-point analysis}
\label{app:placeholders}

In this appendix, we collect the auxiliary functions introduced in Sec.~\ref{subsubsec:AFM-to-PM}.
\begin{widetext}
\begin{align}
    f(d,a)&=\frac{d}{-4+d-2(-6+d)a+2(-6+d)a^2}\,, \label{eq:placeholder-f} \displaybreak[0]\\
    g(N_\phi,d,a)&=\frac{4(-8+d)+(-6+d)dN_\phi}{4[-12+3d-36(a-1)a+7d(a-1)a]+N_\phi[-12+3d-36(a-1)a+8d(a-1)a]}\, \label{eq:placeholder-g}  \displaybreak[0] \\
    h(N_\phi,d)&=\frac{N_\phi+8}{(-6+d)N_\phi+2(-12+d)}\label{eq:placeholder-h}\,, \displaybreak[0]\\
    j(N_\phi,d,a) & = \frac{4d(d-4)-3dN_\phi(d-6)}{k(N_\phi,d)\{4(4-d)[1+3a(a-1)] + N_\phi[-28+7d-84a(a-1) + 18da(a-1)] \}} \,, \displaybreak[0]\\
    k(N_\phi,d) &= 4-d+N_\phi(d-7)\,, \displaybreak[0]\\
    m(N_\phi,d) &= 8(4-d)+3(d-8)N_\phi \,, \displaybreak[0]\\
    p(N_\phi,d)&=-32+(-8+d)N_\phi\label{eq:placeholder-p}\,, \displaybreak[0]\\
    t(d,a)&=\frac{d(d-6)}{(d-10)(-20+5d-60a(a-1)+12da(a-1))}\,, \displaybreak[0]\\
    w(N_\phi,d) &= \frac{6(4-d)N_\phi}{[4-d+(d-7)N_\phi]^2}\,.
\end{align}    
\end{widetext}

\section{Effective Ising field theory for the nematic-to-antiferromagnetic transition (\texorpdfstring{$r_T\rightarrow-\infty$}{rT->-infty})}
\label{app:rt-minus-infty}

In this appendix, we provide further details on the mapping of the full model in Eq.~\eqref{eq:full-Hamiltonian} to an Ising field theory in the limit $r_T\rightarrow-\infty$, relevant for the nematic-to-antiferromagnetic transition discussed in Sec.~\ref{subsubsec:nem-to-AFM}.
The effective interaction in Eq.~\eqref{eq:rt-minus-infty-interaction} is quadratic in the field $\boldsymbol{\phi}$ and can be viewed as an off-diagonal contribution to the mass matrix of the vector field,
\begin{align}
\label{eq:effective-quadratic-term}
\mathcal{H}_{\phi\psi}^{\mathrm{eff}} + \frac{r_\phi}{2} \int \rmd^d x \boldsymbol{\phi}^\top \boldsymbol{\phi}
= \frac{1}{2} \int \rmd^d x \boldsymbol{\phi}^\top M \boldsymbol{\phi}\,.
\end{align}
The corresponding mass matrix is
\begin{equation}
\label{eq:effective-quadratic-matrix}
M=r_\phi\mathds{1}_3 + \frac{\lambda \psi_0}{\sqrt{3}} \sum_{a=1}^3 \eta_a\Lambda_a\,,
\end{equation}
where $\mathds{1}_3$ denotes the $3\times3$ identity matrix and $\Lambda_a$ are the three off-diagonal real Gell-Mann matrices defined in Eq.~\eqref{eq:real-gell-mann}. The matrix $M$ is real and symmetric and can therefore be diagonalized by an orthogonal matrix $R$ according to $M'=RMR^\top$, with
\begin{equation}
\label{eq:rotational-matrix}
R =
\begin{pmatrix}
-\frac{2}{\sqrt{6}}\eta_1\eta_3 & 0 & \frac{1}{\sqrt{3}}\eta_1 \eta_2 \\
\frac{1}{\sqrt{6}}\eta_2\eta_3  & \frac{1}{\sqrt{3}}\eta_1 \eta_3 & \frac{1}{\sqrt{3}}\eta_2 \eta_3 \\
\frac{1}{\sqrt{6}} & -\frac{1}{\sqrt{2}} \eta_1\eta_2 & \frac{1}{\sqrt{3}}
\end{pmatrix}\,.
\end{equation}
Defining the rotated field as $\boldsymbol{\phi}'=R^\top\boldsymbol{\phi}$, the vector-field Hamiltonian in Eq.~\eqref{eq:vector-part} becomes
\begin{equation}
    \mathcal{H}^{\mathrm{eff}}_\phi=\int \rmd^dx \,\left[\frac{1}{2}\sum_{i}(\nabla\phi_i')^2+ \frac{1}{2}{\boldsymbol{\phi}'}^{\top} M'\boldsymbol{\phi}' + \mathcal O({\boldsymbol{\phi}'}^{4}) \right]
\end{equation}
where the mass matrix is diagonal, $M'=\diag(m_+,m_+,m_-)$, with eigenvalues
\begin{align}
\label{eq:new-masses}
m_+ & =r_\phi-\frac{\lambda\psi_0}{\sqrt{3}}\eta_1\eta_2\eta_3\,, &
m_- & =r_\phi+\frac{2\lambda\psi_0}{\sqrt{3}}\eta_1\eta_2\eta_3\,.
\end{align}
For our choice $\omega,\lambda>0$, the nematic order parameter aligns along one of the four cubic diagonals shown in Fig.~\ref{fig:minima-sketches}(b), for which $\eta_1\eta_2\eta_3=-1$. Consequently, $m_-<m_+$, and the $\phi_3'$ mode becomes critical first as $r_\phi$ is reduced. The modes $\phi_1'$ and $\phi_2'$ remain gapped at this transition and can therefore be neglected when determining the critical behavior. This yields an effective theory for the $\phi_3'$ mode alone,
\begin{equation}
\label{eq:effective-Z2}
\mathcal{H}^{\mathrm{eff}}_{\phi_3'}=\int \rmd^d x \left[\frac{1}{2}(\nabla\phi_3')^2+\frac{m_-}{2}{\phi_3'}^2+\frac{3u_\phi+v_\phi}{3 \cdot 4!} {\phi_3'}^4 \right]\,,
\end{equation}
which is the conventional Ising field theory with $\mathbb{Z}_2$ symmetry, as stated in the main text. In the rotated frame, this effective theory is independent of the particular $\eta_a$ configuration, provided the constraint $\eta_1\eta_2\eta_3=-1$ is satisfied.

\bibliographystyle{longapsrev4-2}
\bibliography{nem-FT}

\end{document}